\newcommand{\PaperTitle}{ Network Function  Capacity Reconnaissance  by Remote Adversaries}
\newcommand{\PaperNumber}{665}

\documentclass[10pt,sigconf,letterpaper,nonacm]{acmart}

\usepackage{xspace}
\usepackage{subcaption}
\usepackage{multirow}
\usepackage[labelfont=bf,font=small,skip=2pt,belowskip=-1pt]{caption}
\usepackage{pifont}
\usepackage{graphicx}
\usepackage{tabularx}

\begin{document}

\title{\PaperTitle}
\author{Aqsa Kashaf}
\email{aqsa.kashaf@gmail.com}
\affiliation{
\institution{Bytedance Inc.}
\country{USA}
}
\author{Aidan Walsh}
\email{abwalsh@princeton.edu}
\affiliation{%
  \institution{Princeton University}
  \country{USA}
}
\author{Maria Apostolaki}
\email{abwalsh@princeton.edu}
\affiliation{%
  \institution{Princeton University}
  \country{USA}
}
\author{Vyas Sekar}
\email{vsekar@andrew.cmu.edu}
\affiliation{%
  \institution{Carnegie Mellon University}
  \country{USA}
}
\author{Yuvraj Agarwal}
\email{yuvraj@cs.cmu.edu}
\affiliation{%
  \institution{Carnegie Mellon University}
  \country{USA}
}

\newcommand{\Section}{\S}
\newcommand{\highlight}[1]{\textcolor{black}{#1}}
\newcommand{\norm}[1]{\left\lVert#1\right\rVert}
\newcommand{\parentheses}[1]{\left(#1\right)}
\newcommand{\brackets}[1]{\left[#1\right]}
\newcommand{\set}[1]{\left\{#1\right\}}
\newcommand{\abs}[1]{\left|#1\right|}
\newcommand{\indicator}{\ensuremath{\mathbbm{1}}}

\newcommand{\ie}{\emph{i.e.,}\xspace}
\newcommand{\eg}{\emph{e.g.,}\xspace}
\newcommand{\etc}{\emph{etc.}\xspace}
\newcommand{\etal}{\emph{et al.}\xspace}
\newcommand{\rmr}[1]{\st {#1}}
\newcommand{\sysname}{{\sys}\xspace}

\newcommand{\name}{\sysname}
\newcommand{\dnf}{\textsc{Delay\textunderscore NF}\xspace}
\newcommand{\surbl}{\textsc{SUR-BL}\xspace}
\newcommand{\surrl}{\textsc{SUR-RL}\xspace}
\newcommand{\surblmulti}{SUR-BL-mt\xspace}
\newcommand{\surrlmulti}{SUR-RL-mt\xspace}
\newcommand{\snortrl}{\textsc{SNORT-RL}\xspace}
\newcommand{\slops}{\textit{SLoPS}\xspace}

\newcommand{\remove}[1]{}

\newcommand{\myitem}[1]{\noindent\textbf{#1}}

\newlength\mylen
\newcommand\myinput[1]{%
  \settowidth\mylen{\KwIn{}}%
  \setlength\hangindent{\mylen}%
  \hspace*{\mylen}#1\\}
\newcommand{\dvfslength}{5K\xspace}  
\newcommand{\sys}{\textit{NFTY}\xspace}
\newcommand{\syslite}{\textit{NFTY-100}\xspace}
\newcommand{\syslarge}{\textit{NFTY-\dvfslength}\xspace}
\newcommand{\trainsize}{\ensuremath{L}\xspace}
\newcommand{\disp}{\ensuremath{\delta}\xspace}
\newcommand{\ithdisp}[1]{\ensuremath{\delta_{#1}}\xspace}
\newcommand{\bone}{mean-train\xspace}
\newcommand{\btwo}{median-train\xspace}
\newcommand{\bthree}{mdn\_disp\xspace}

\newcommand{\greentick}{{\color{green}\ding{51}}} 
\newcommand{\redcross}{{\color{red}\ding{55}}} 

\newcommand{\FFunction}{\ensuremath{\mathit{F}}}
\newcommand{\NF}{\textit{NF}\xspace}
\newcommand{\NFCR}{\textit{NFCR}\xspace}
\newcommand{\NFs}{\textit{NFs}\xspace}
\newcommand{\NFFull}{network function\xspace}
\newcommand{\stepDetectionAlgo}{Binary Segmentation\xspace}
\newcommand{\NFType}{\ensuremath{\mathit{NFType}}}
\newcommand{\NFTypeIndex}{\ensuremath{\mathit{t}}}
\newcommand{\NFInstance}{\ensuremath{\mathit{NFInst}}}
\newcommand{\NFInstanceIndex}{\ensuremath{\mathit{i}}}
\newcommand{\CMIndex}{\ensuremath{\mathit{c}}}
\newcommand{\NumNFInstances}{\ensuremath{\mathit{NumNF}}}
\newcommand{\NumArchs}{\ensuremath{\mathit{NumArch}}}
\newcommand{\ContBefore}{\ensuremath{\mathit{Cont\_before}}}
\newcommand{\ContAfter}{\ensuremath{\mathit{Cont\_after}}}
\newcommand{\Config}{\ensuremath{\mathit{Conf}}}
\newcommand{\Traffic}{\ensuremath{\mathit{Traffic}}}
\newcommand{\Arch}{\ensuremath{\mathit{Arch}}}
\newcommand{\ArchIndex}{\ensuremath{\mathit{k}}}
\newcommand{\Competition}{\ensuremath{\mathit{Comp}}}
\newcommand{\CompetitionIndex}{\ensuremath{\mathit{j}}}
\newcommand{\MaxNumContenders}{\ensuremath{\mathit{N}}}
\newcommand{\NumContenders}{\ensuremath{\mathit{n}}}
\newcommand{\TrafficRate}{\ensuremath{\mathit{ThruPut}}}

\newcommand{\Performance}{\ensuremath{\mathit{Perf}}}
\newcommand{\Predicted}{\ensuremath{\mathit{Pred}}}
\newcommand{\Observed}{\ensuremath{\mathit{Obs}}}

\newcommand{\Sensitivity}{\ensuremath{\mathit{Sens}}}
\newcommand{\Bubble}{\ensuremath{\mathit{Bubble}}}
\newcommand{\Synth}{\ensuremath{\mathit{Synth}}}

\newcommand{\BubbleConf}{\ensuremath{\mathit{WSS}}}
\newcommand{\MinBubble}{\ensuremath{\mathit{WSSmin}}}
\newcommand{\MaxBubble}{\ensuremath{\mathit{WSSmax}}}

\newcommand{\RReqs}{\ensuremath{\mathit{RR}}}
\newcommand{\MinRReqs}{\ensuremath{\mathit{RRmin}}}
\newcommand{\MaxRReqs}{\ensuremath{\mathit{RRmax}}}

\newcommand{\Reporter}{\ensuremath{\mathit{Reporter}}}
\newcommand{\Contentiousness}{\ensuremath{\mathit{Cont}}}
\newcommand{\PerfModel}{\ensuremath{\mathit{Model}}}

\newcommand{\Perf}{\ensuremath{\mathit{Perf}}}
\newcommand{\Dataset}{\ensuremath{\mathit{D}}}
\newcommand{\MemBW}{\ensuremath{\mathit{MemBW}}}

\newcommand{\ArchFootprint}{\ensuremath{\mathit{CM}}}
\newcommand{\ArchFootprintIndex}{\ensuremath{\mathit{l}}}

\newcommand{\aqsa}[1]{\textcolor{blue}{[aqsa: #1]}}
\newcommand{\new}[1]{\textcolor{blue}{[#1]}}
\newcommand{\red}[1]{\textcolor{red}{#1}}
\newcommand{\maria}[1]{{\color{orange} {(maria: #1)}}}
\newcommand{\YA}[1]{{\color{blue} {(YA: #1)}}}
\newcommand{\vyas}[1]{{\color{red} {(vyas: #1)}}}
\newcommand{\mycircle}[1]{\mbox{\textcircled{#1}}}

\newcommand{\cp}{\ensuremath{\mathbb{C}_p}\xspace}
\newcommand{\pkttype}{\ensuremath{p^t}\xspace}
\newcommand{\budget}{\ensuremath{\mathbb{B}}\xspace}

\newcommand{\probe}{\ensuremath{\rho}\xspace}

\newcommand{\syslargeerrorlab}{9\%\xspace}
\newcommand{\sysonesidederror}{10\%\xspace}
\newcommand{\syslargedelta}{30x\xspace}
\newcommand{\sysliteerrorlab}{9\%\xspace}
\newcommand{\syslitedelta}{20x\xspace}
\newcommand{\sysliteaws}{7\%\xspace}
\newcommand{\stepdetectiondelta}{25x\xspace}
\newcommand{\measurementoptdelta}{33x\xspace}
\newcommand{\syslargeinterneterror}{9\%\xspace}
\newcommand{\sysliteinterneterror}{10\%\xspace}

\newcounter{packednmbr}

\newcommand{\mypara}[1]{\noindent{\bf {#1}:}~}
\newcommand{\myparait}[1]{\noindent{\em {#1}:}~}
\newcommand{\myparatight}[1]{\noindent{\bf {#1}:}~}
\newcommand{\myparaq}[1]{\medskip\noindent{\bf {#1}?}~}
\newcommand{\myparaqtight}[1]{\noindent{\bf {#1}?}~}
\newenvironment{packedenumerate}{\begin{list}{\thepackednmbr.}{\usecounter{packednmbr}\setlength{\itemsep}{0.3pt}\addtolength{\labelwidth}{-4pt}\setlength{\leftmargin}{\labelwidth}\setlength{\listparindent}{\parindent}\setlength{\parsep}{0.5pt}\setlength{\topsep}{0pt}}}{\end{list}}

\newenvironment{packeditemize}{\begin{list}{$\bullet$}{\setlength{\itemsep}{0.3pt}\addtolength{\labelwidth}{-4pt}\setlength{\leftmargin}{\labelwidth}\setlength{\listparindent}{\parindent}\setlength{\parsep}{0.5pt}\setlength{\topsep}{0pt}}}{\end{list}}
\newenvironment{packedtrivlist}{\begin{list}{\setlength{\itemsep}{0.3pt}\addtolength{\labelwidth}{-4pt}\setlength{\leftmargin}{\labelwidth}\setlength{\listparindent}{\parindent}\setlength{\parsep}{0.5pt}\setlength{\topsep}{0pt}}}{\end{list}}



\newcommand{\insightref}[1]{\textbf{Obs~}\ref{#1}}
\newcounter{insightlabel}
\newcounter{insightnmbr}
\renewcommand{\theinsightlabel}{\textbf{\theinsightnmbr}}
\newenvironment{insight}{
\begin{list}{\textbf{Observation }\theinsightlabel:~}{\usecounter{insightlabel}\stepcounter{insightnmbr}\setlength{\labelwidth}{0pt}\setlength{\labelsep}{0pt}\setlength{\leftmargin}{0in}\noindent\rule{\linewidth}{1pt}\vspace{-2pt}\item \bf \em}}{\\[-7pt]\end{list}\vspace{-10pt}\noindent\rule{\linewidth}{1pt}}


\begin{abstract}

There is anecdotal evidence that attackers use reconnaissance to learn the capacity of their victims before DDoS attacks to maximize their impact. The first step to mitigate capacity reconnaissance attacks is to understand their feasibility. However, the feasibility of capacity reconnaissance in network functions (\NFs) (e.g., firewalls, NATs) is unknown. To this end, we formulate the problem of network function capacity reconnaissance (\NFCR) and explore the feasibility of inferring the processing capacity of an NF while avoiding detection.
We identify key factors that make \NFCR challenging and analyze how these factors affect accuracy (measured as a divergence from ground truth) and stealthiness (measured in packets sent). We propose a flexible tool, NFTY, that performs NFCR and we evaluate two practical NFTY configurations to showcase the stealthiness vs. accuracy tradeoffs. We evaluate these strategies in controlled, Internet and/or cloud settings with commercial NFs.
\sys can accurately estimate the capacity of different \NF deployments within \sysliteinterneterror error in the controlled experiments and the Internet, and within  \sysliteaws error for a commercial \NF deployed in the cloud (AWS). Moreover, \sys outperforms link-bandwidth estimation baselines by up to \syslargedelta. 
\end{abstract}
\settopmatter{printfolios=true}
\maketitle

\section{Introduction}

Attackers seek to optimize their impact by using reconnaissance probes to assess a target's resources before launching an attack. For example, prior work has shown attackers can use topology information to launch sophisticated denial-of-service (DoS) attacks and concentrate their efforts on important nodes or links of a targeted network~\cite{panjwani2005experimental,kang2013crossfire}. 
Particularly in attacks against critical network infrastructures, a significant aspect of reconnaissance involves estimating the capacity of network functions (NFs), which implement crucial operations, including DDoS mitigation, firewall management, and network translation.
Anecdotal evidence suggests that attackers might be conducting such reconnaissance to estimate the capacity of NFs before initiating DDoS attacks (e.g., ~\cite{schneier_2016,verisign_blog_2016,corero}). Accurately gauging the capacity of various \NFFull enables an adversary to identify the most vulnerable targets and estimate the necessary resources (e.g., number of packets, attack rate) for a successful attack.

Despite the potential risks and impact of such attacks, little is known about the feasibility of Network Function Capacity Reconnaissance (\NFCR).
This paper aims to raise awareness of \NFCR  attacks and guide potential targets towards practical countermeasures. 
To the best of our knowledge, we are the first to comprehensively define the problem of \NFCR.
Specifically, we are interested in answering: \emph{Can an attacker infer the processing capacity of an \NF remotely while remaining undetected (\eg without sending too many packets)?}

At first glance, it seems an attacker can leverage the significant prior work on link capacity estimation~\cite{jain2002end,lai2001nettimer,carter1996measuring,kapoor2004capprobe} for \NFCR. 
Yet, we find that link-capacity estimation techniques are ineffective in the \NFCR context due to several unique challenges.
First, unlike link-capacity estimation techniques, which often involve sending large volumes of traffic\cite{tirumala1999iperf,mathis1996diagnosing}, the number of packets sent for \NFCR is crucial for its success. 
Intuitively, sending more traffic increases the likelihood of the target detecting the impending attack and bolstering capacity. Second, most link capacity estimation techniques~\cite{saroiu2002sprobe,carter1996measuring} require control over two vantage points before and after the link of interest, which is not always easy to achieve, as \NFs can be deployed within private networks. 
Third, \NFCR is affected by optimizations in the NF deployment such as multi-threading and CPU frequency scaling, which are not typical to network links whose capacity is stable, hence the bandwidth estimation techniques~\cite{bolot1993characterizing,carter1996measuring,dovrolis2001packet,jain2002end,kapoor2004capprobe,lai1999measuring,paxson1997end} are mostly ineffective in \NFCR.

Despite the challenges, we demonstrate the feasibility of \NFCR by designing and  implementing
a functional attacker strategy, which we call \NF capaciTY estimation
(\sys) that would allow an attacker to remotely and accurately estimate an \NF's capacity while sending a relatively small number of packets to evade detection.
At a high level, \sys operates by sending a burst of packets designed to momentarily stress the NF, revealing its true capacity despite hardware and software optimizations (not relevant in link capacity estimation). 
\sys measures the dispersion of the packets in the burst —a concept borrowed from link-capacity estimation techniques— but innovatively processes the dispersion measurements to reduce noise.  \sys is not only much more accurate than link-capacity estimation baselines (up to \syslargedelta) but also generalizes to an attacker controlling only a single vantage point by leveraging the TTL-exceeded mechanism in Internet routers and surpassing other associated practical issues (\eg ICMP rate-limiting, router bottlenecks). 

We design \sys to be configurable depending on the attacker's capability (\eg vantage points at sender and receiver), and her knowledge of the target \NF's deployment (if any).
Among all possible \sys configurations, we 
identify two concrete representative strategies and evaluate them in lab settings and in the wild (public Internet, Internet2, cloud). The two attacks, namely \syslite and \syslarge, differ in the number of packets sent and thus their effectiveness and visibility. At a high level, \syslite is stealthier but less accurate. Specifically, \syslite works for an attacker that controls two vantage points and sends only $100$ packets from one to the other. \syslite can accurately estimate the capacity of simple NFs with \sysliteerrorlab in the controlled environment and within \sysliteinterneterror error in the Internet.
We have used \syslite to estimate the capacity of a commercial NF in AWS within \sysliteaws of its true capacity while avoiding detection. 

\syslarge, in contrast, works even with an attacker controlling a single vantage point and against more complex NFs but is less stealthy as it sends \dvfslength packets. We find that \syslarge estimates a more diverse set of NFs with up to \syslargeerrorlab error in the controlled environment and within \syslargeinterneterror error in the Internet. This is up to \syslargedelta more accurately than the bandwidth estimation baselines.
 
We also propose practical countermeasures against \NFCR attacks. Specifically, we consider countermeasures that aim at adding noise to probing traffic or obfuscating their infrastructure with rate-limiting and evaluating their effectiveness and performance overhead.

\mypara{Contributions}
\begin{packeditemize}
    \item We are the first to formally define and analyze the feasibility of  \NF Capacity Reconnaissance (\NFCR). 
   \item We investigate the applicability of bandwidth estimation techniques to \NFCR and identify the unique challenges that differentiate \NF capacity estimation from bandwidth estimation.
    \item We introduce \sys, a flexible methodology for conducting \NFCR, that is accurate, stealthy, and practical even for an attacker that is single-sided (\ie cannot measure the packet it sends directly).
    \item We evaluate \sys on both lab and Internet settings as well as against a commercial NF in the cloud.
    \item We present a comprehensive suite of countermeasures designed to safeguard potential targets against \NFCR  attacks with limited overhead.
\end{packeditemize}
\section{Problem Formulation} 
\label{sec:problemDef}
In this section, we introduce the general problem of \NF Capacity Reconnaissance (\NFCR) using a representative network setting. 
Next, we specify the threat model and scope our problem.   

\label{ssub:setting}
\begin{figure}[t!]
    \centering
    \includegraphics[width=0.9\columnwidth]{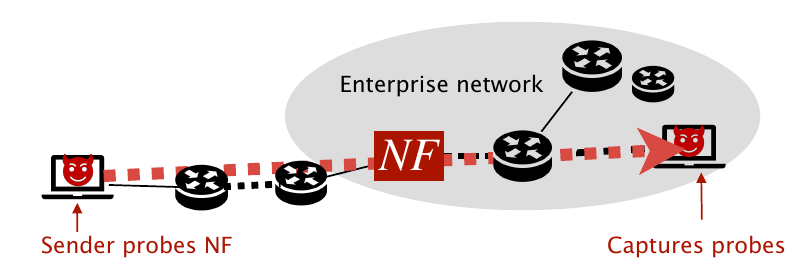}
    \caption{\small The \NF processes incoming traffic of the private enterprise. A two-sided attacker controls nodes on both sides of the \NF (in and out of the enterprise), while a one-sided attacker controls nodes only outside the enterprise (sender).\label{fig:setting}}
    \Description[Paper Setting]{The \NF processes the incoming traffic of the private enterprise. A two-sided attacker controls nodes on both sides of the \NF (in and out of the enterprise), while a one-sided attacker controls nodes only outside the enterprise.}
    \vspace{-4pt}
\end{figure}

\myparatight{Setting} 
We assume an enterprise that deploys a \NFFull (\NF) (e.g. IPS, firewall) placed such that it observes all incoming traffic, as shown in Figure~\ref{fig:setting}. The \NF comprises a set of processing tasks applied to a packet type $t$ with a well-defined goal, such as load balancing or intrusion detection. 
For example, a TCP SYN proxy NF is geared towards protecting against \textit{TCP SYN} flooding attacks. An \NF can be deployed on different hardware (\eg a server, a programmable switch) with resources configured statically or dynamically (\eg number of threads, cores, memory). 
 We define capacity \cp of an \NF as the maximum number of packets of type t (\pkttype) the \NF can process per unit time.
An \NF can have different capacities for different packet types.
Note that \NFCR is not specific to a private enterprise setting. It generalizes to other settings, such as the cloud, in which the \NFs are deployed by various third parties \eg the Azure Marketplace~\cite{azure-marketplace}.

\myitem{Threat Model:} We consider two threat models depending on the attacker's capabilities as either \emph{(i)} one-sided or \emph{(ii)} two-sided; depicted in Figure~\ref{fig:setting}. In the two-sided threat model, the attacker controls a vantage point (VP) on both sides of the \NF. The attacker can send traffic from one VP (e.g., a host on the Internet) to the other (a host inside the enterprise network) via the \NF. In the one-sided threat model, the attacker controls a single VP (e.g., a random Internet host).  
In both models, the attacker is unaware of the deployment of the \NF; the topology of the target network and location of the \NF; the congestion on the network path to the \NF, and the \NF itself. The attacker aims to have a low network footprint to avoid detection. Beyond preventing being blocked, the attacker's low footprint prevents the victim from noticing the reconnaissance and ramping up resources in preparation for the real attack.
Intuitively, the more packets an attacker sends, the more visible and hence detectable they will be.

\myitem{Problem Definition and Scope:}
Given this context, we formally define the \NFCR problem as follows:
 Given an \NF  with unknown processing capacity \cp for packet type \pkttype, unknown location, and deployment, construct an inference model that can remotely and accurately estimate the capacity \cp for packet \pkttype by probing it with at most budget \budget packets. 





In this paper,  we restrict our focus to NFs whose processing capacity does not depend on the packet history. The NF may maintain state, but that does not affect its processing capacity, \ie an attacker cannot decrease the processing capacity of the target NF by sending certain packet sequences.
\footnote{Most Internet-facing NFs have this property to prevent an attacker from algorithmic attacks exploiting this dependency to deplete the NF's resources. }
We also only consider statically provisioned \NFs, meaning that their allotted resources (cores, memory, hardware) do not change with time. Observe that the NFs' deployment can still use multi-threading or CPU optimizations but are statically provisioned.
This is common in practice, especially for small/medium enterprises, as dynamically provisioning an \NF is complex.\footnote{
Automating the entire lifecycle of NFs, from provisioning to decommissioning, while handling workload fluctuations, real-time performance monitoring, and decision-making makes dynamic provisioning a highly complex and nuanced task.}
We consider an attacker that knows the high-level packet types that the target \NF processes (\eg TCP SYN, UDP). 
The measured capacity will correspond only to this packet type.
This is not a limitation given that the attacker can typically use packet types that correspond to commonly seen traffic, e.g., TCP SYN, HTTP, DNS, and observe whether the output changes to identify which are processed.
Finally, similarly to bandwidth-estimation techniques~\cite{bai1997estimating,bolot1993characterizing,clink,dovrolis2001packet,jacobson1997pathchar,jain2002end, johnsson2004diettopp} which assume that the target link is the bottleneck, \NFCR assumes that the \NF is the bottleneck in the path (\ie its processing rate is lower than that of the links).


\section{Prior work and limitations}

A natural starting point for an attacker is to use link-bandwidth estimation techniques for \NFCR. 
Hence, in this section, we try such techniques against real NFs and show that they do not provide an accurate capacity estimate. 
Finally, we reveal the unique characteristics of \NFCR that make traditional link-bandwidth estimation techniques ineffective.

\subsection{Overview of Bandwidth Estimation}
Based on the probing technique and the signal used, bandwidth estimation techniques can be categorized into:  

\myitem{Dispersion-based} techniques typically send a sequence or pair of packets and measure the link-induced spacing at the receiver~\cite{carter1996measuring,jacobson1997pathchar,li2008wbest,dovrolis2004packet}. 
\myitem{Rate-based techniques} iteratively probe at varying rates until they find the minimum rate that causes the one-way delay of packets to increase, signaling self-induced congestion~\cite{ribeiro2003pathchirp,jain2002end}. 
\myitem{Bulk-transfer} techniques send a high traffic volume for a longer duration and measure throughput (\ie after self-induced drops) at the receiver~\cite{tirumala1999iperf,allman2001measuring}. Using bulk transfer, an attacker might try to flood the \NF by sending packets at a high rate until drops are observed.

\subsection{An Experimental Insight}\label{sec:challenges}

\begin{figure}[t!]
    \centering
    \includegraphics[width=0.75\columnwidth]{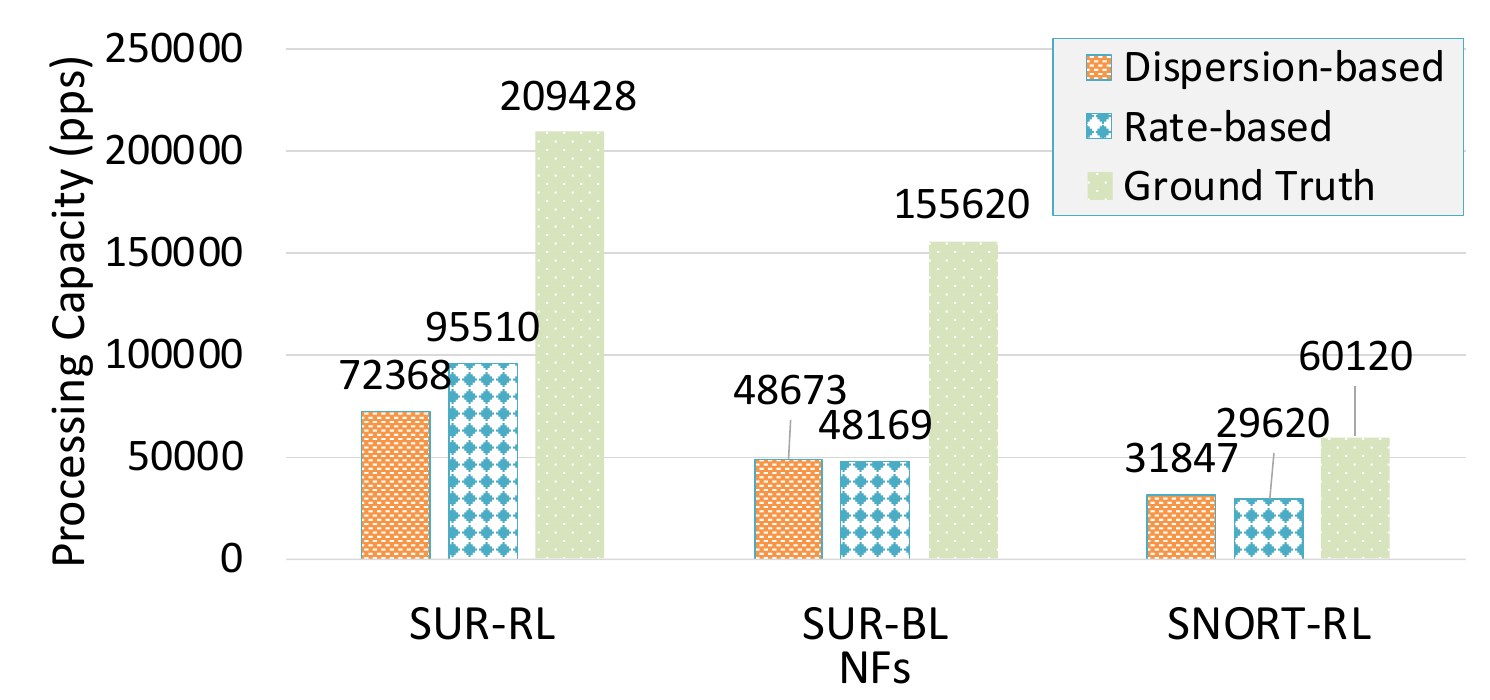}
    \caption{\small Traditional bandwidth-estimation techniques underestimate the processing capacity of \NFs.
    \Description[Traditional bandwidth estimation techniques]{Traditional bandwidth-estimation techniques underestimate the processing capacity of \NFs.}
    \label{fig:fs-ex}}
    \vspace{-4pt}
\end{figure}

We start by using each of these techniques to measure the capacity of realistic NFs built on top of Snort~\cite{SnortNetworkIntrusion} and Suricata~\cite{Suricata} and illustrate the results in Figure~\ref{fig:fs-ex}.
We observe a huge discrepancy among the three approaches. 
Naturally, the bulk transfer is the most accurate.
By offering excessive load, the NF runs at maximum capacity; thus, the measured throughput is the capacity. 
While bulk transfer can serve as an experimental ground truth, it does not satisfy \NFCR. As bulk transfer requires sending excessive traffic for an extended period to induce drops, it will be extremely noticeable.
Unfortunately, the less noticeable dispersion-based and rate-based techniques (orange and blue bars) are extremely inaccurate, with up to 65\% and 54\% error, respectively.

\subsection{Challenges of \NFCR }
We describe the characteristics that make \NFCR challenging and traditional bandwidth estimation techniques ineffective. 

\myitem{Optimized NF Deployments make NF capacity dynamic.} 
Although we have scoped the problem to statically provisioned \NF (in \S\ref{sec:problemDef}), software and hardware optimizations (\eg CPU multi-threading or frequency scaling) still introduce a level of variability in an NF's effective capacity, which can change based on current load conditions or the types of packets being processed. This variability presents a significant challenge for Network Function Capacity Reconnaissance (NFCR) techniques, which might underestimate the NF's capacity. 
In the context of bandwidth estimation, such variability does not arise as there is no optimization affecting link capacity.
In the NF context, though,  multi-threading --often used to increase the CPU efficiency-- can change the measured aggregate processing capacity of an NF.
Similarly, Dynamic Voltage and Frequency Scaling (DVFS) -- used to adjust CPU frequency based on load to reduce power consumption-- can cause the attacker to underestimate the capacity. 
Finally, unlike links for which same-length packets are processed at a similar speed, NF's speed might be affected by the packet type or flow distribution, \eg due to thread scheduling, hence there is need for more diverse probes.




\myitem{Two-sided estimation techniques do not generalize to one-sided.} Traditional bandwidth-estimation techniques typically assume two-sided control of the link. This is natural as those techniques were designed as a benign tool for operators controling their network. In this paper, we aim to investigate the effectiveness of an attacker performing \NFCR that might only have one-sided control. Link bandwidth estimation techniques that work under the one-sided assumption typically rely on a particular protocol behavior involving specific packet types (\eg a TCP SYN on a closed port triggers a TCP RST~\cite{saroiu2002sprobe}) and thus do not generalize to any \pkttype that the target NF is processing. 

\myitem{Measurement infrastructure introduces noise.}
Intuiti\-vely, the used infrastructure is crucial for any measurement campaign. 
In the context of \NFCR, accuracy in sending packets at a concrete rate and monitoring packets' arrival timestamps is crucial and surprisingly hard to achieve in practice. 
Intuitively, optimization at the receiver, \eg batching in software (NAPI~\cite{salim2001beyond}) or hardware (IC~\cite{IC}), result in very coarse-grained packet timestamps, which might undermine the measurements. Similarly, software-based packet generation is easier to implement but is unlikely to offer reliable rates for probing. 
While these are concerns shared with bandwidth-estimation, we report our unique insights in \S\ref{sec:measurements}. 
\section{Design}

We begin by explaining the main insights that drive \sys's design(\S\ref{sec:overview}). Then, we elaborate on how we adapt dispersion-based capacity estimation techniques to be effective in \NFCR (\S\ref{sec:dispersion}). Finally, we discuss how we can adapt \sys to cases where the attacker controls only one measurement endpoint (\S\ref{sec:onesided}). 
Section~\ref{sec:e2e} presents an end-to-end view of how one can configure and use the \sys tool in practice.


\subsection{ High-level Overview \& Insights} \label{sec:overview}
\begin{figure}[t!]
    \centering
    \includegraphics[width=0.9\columnwidth]{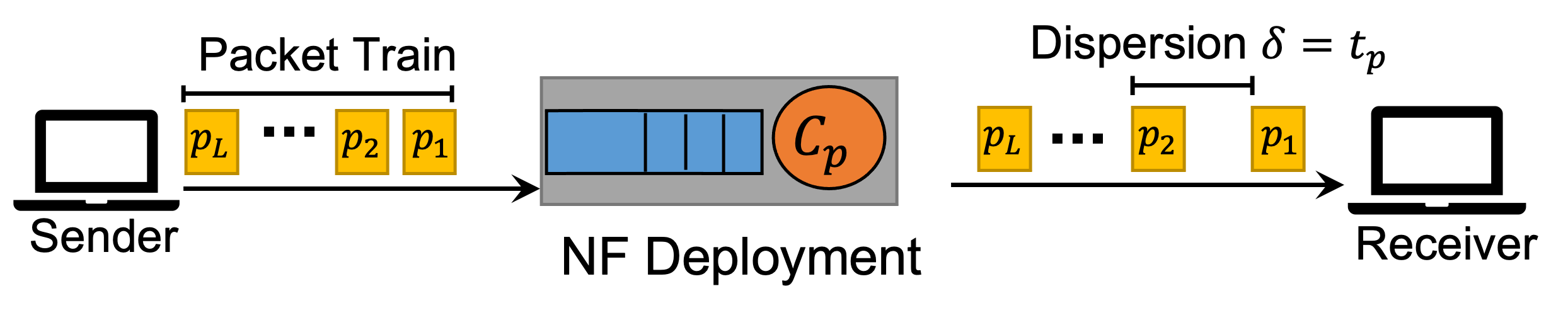}
    \vspace{-0.1cm}
    \caption{\small In the two-sided threat model, the time difference between consecutive packets increases as they exit from the \NF, revealing \NF's processing time ($\delta=t_p$), thus its capacity. }
    \label{fig:disp}
    \Description[two-sided threat model]{In the two-sided threat model, the time difference between consecutive packets increases as packets exit from the \NF, revealing \NF's processing time ($\delta=t_p$), thus its capacity. }
    \vspace{-0.4cm}
\end{figure}

Consider an attacker who aims to estimate the maximum packet processing rate of an NF (deployed at the edge of a victim network as in Figure~\ref{fig:setting}) for a particular packet type (\eg UDP). 
Dispersion-based techniques are the starting point for \sys, as they offer a stealthier approach by sending fewer packets. Following that strategy, the attacker sends a set of UDP packets to an attacker-controlled receiver via the target NF. 
The receiver measures the dispersion \ie the NF-induced time gap between consecutive packets, to infer the NF processing capacity. After being processed by the NF, the packets will be spaced in time depending on the NF processing time (Figure~\ref{fig:disp}).

Our first insight is that to achieve high accuracy, the attacker needs to trigger the NF's highest capacity during her measurements. To do so, while remaining stealthy, the attacker needs to send bursts of packets, which will instantaneously stress the \NF with the minimum number of packets. Furthermore, the attacker collects measurements at the receiver and needs to process them to identify step-wise patterns that reveal capacity adaptations triggered by the NF's host (\S \ref{sec:dispersion}). 
Our second insight extends \sys to an attacker not controlling the receiver, \ie one-sided threat model. Concretely, we find that routers on-path can inadvertently act as malicious receivers (reporting dispersion values) if they are triggered to send ICMP error messages back to the attacker-controlled sender. Importantly, any IP packet can trigger an ICMP error message using an appropriate TTL value on any router on-path past the NF. While doing so is not trivial, as routers often rate-limit such requests or delay them, we explain how the attacker can use this insight to accurately estimate NF's capacity with one-sided control in \S \ref{sec:onesided}.
Finally, in both cases, the measurement infrastructure can make (or break) the attack by offering (in)accurate timestamping and (in)accurate probing rates. We elaborate on our measurement methodology in \S \ref{sec:measurements}.




\subsection{Adapting dispersion-based bandwidth estimations to \NFCR}\label{sec:dispersion}
We start with the essential background of dispersion-based estimation and discuss our observations as we attempt to adapt it to \NFCR.  Finally, we describe our approach.

\myitem{Dispersion-based link-capacity estimation techniques} seek to estimate the bandwidth of a bottleneck link by sending a probe of two (i.e., a pair) or more (i.e., a train) packets back to back and measuring their spacing at the receiver. Intuitively, for a packet pair, assuming that the link capacity of the bottleneck link is lower than the packets' inter-arrival rate, the second packet of a packet pair will be queued, waiting for the transmission of the first one. The time difference of the arrival between the two packets (measured at the destination) is called {\em dispersion \disp} and provides the transmission time $t_{tr}$ on the bottleneck link. We can then calculate the bottleneck link capacity $C_l$ for a packet of size $S$ as follows:  $C_l  =  S / t_{tr} = S / \delta$. Bandwidth estimation techniques 
rely on the assumption that processing delays on the path are negligible compared to link transmission delays.

\myitem{Dispersion-based \NFCR} seeks to estimate the capacity of any NF in processing any particular packet type by sending packet pairs or trains.
If the pair/train traverses an \NF, the dispersion gives the \NF's processing time $t_p$, which is typically higher than transmission time since \NF's do more than just forwarding. In the context of \NFCR 
the dispersion is:
\begin{align}
    \delta = \max (t_{tr}, t_p) = \max (S/C_l,1/\cp)
\end{align}
Thus, if the \NF's processing time $t_p$ is higher than the transmission time $t_{tr}$, then the measured dispersion corresponds to the former.



\begin{figure}[t!]
    \centering
    \begin{subfigure}[b]{0.49\columnwidth}
    \small
    \centering
    \includegraphics[width=\columnwidth]{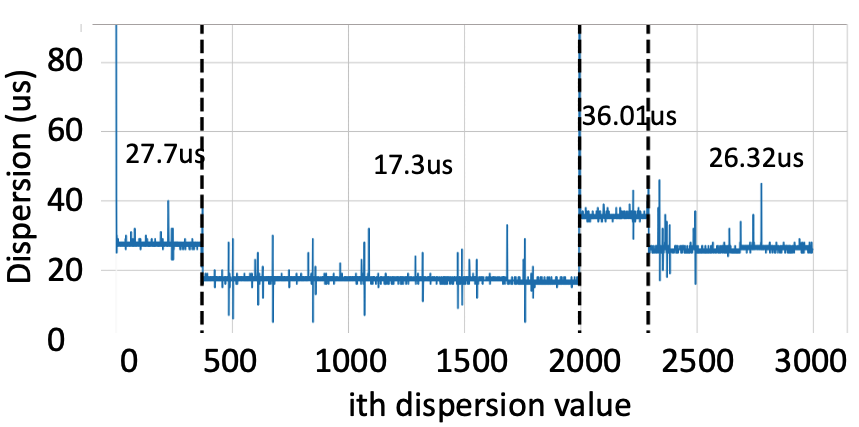}
    \caption{\label{fig:fs2-ex}}
    \end{subfigure}
    \begin{subfigure}[b]{0.49\columnwidth}
    \small
    \centering
    \includegraphics[width=\columnwidth]{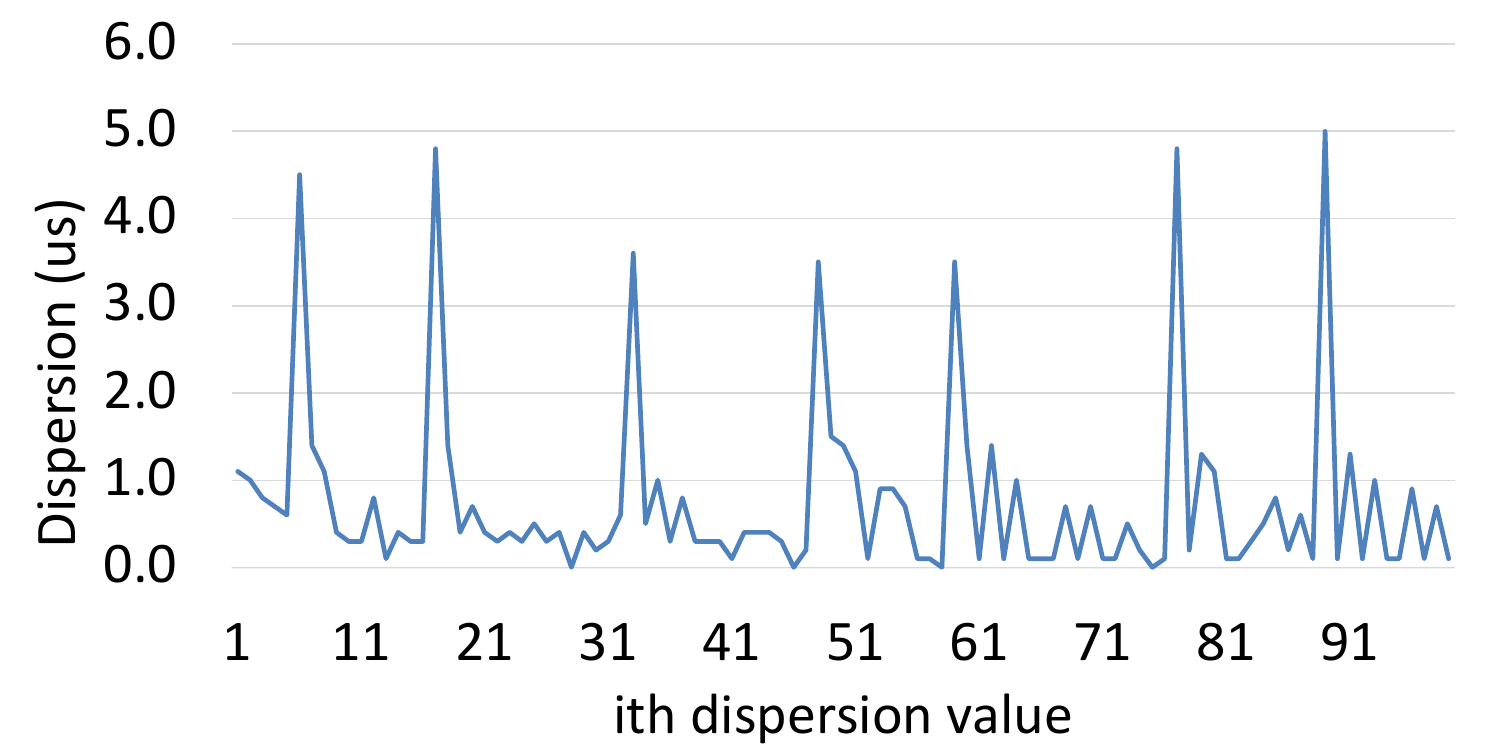}
    \caption{\label{fig:batch-spikes}}
    \end{subfigure}
    \caption{\small Effect of DVFS and packet batching on dispersion values, affecting capacity estimates. (a) Assuming that the last step in dispersion values corresponds to maximum frequency would be incorrect due to the potential under-clocking. (b) Packet batching can cause spikes in the dispersion signature.
    }
    \Description[Effect of DVFS and packet batching on dispersion values, affecting capacity estimates]{Effect of DVFS and packet batching on dispersion values, affecting capacity estimates. (a) Assuming that the last step in dispersion values corresponds to maximum frequency would be incorrect in this example due to the potential under-clocking. (b) Packet batching can cause spikes in the dispersion signature. Hence, choosing a probe length larger than the batch interval is important.}
     
\end{figure}

\myitem{Observation 1: Not all dispersion values correspond to the true capacity.} 
Fig.~\ref{fig:fs2-ex} shows the dispersion values that we actually observed at the \NF.  One cause of these transitions is that the NF runs on a CPU configured with DVFS: the CPU operates at a lower frequency under a lighter load and transitions to a higher frequency under a heavier load driven by the OS or kernel. As a CPU might undergo multiple transitions to different frequencies, the observed rate at which packets are processed will change. As a result, the measured dispersion values may not all correspond to the true capacity, and traditional link bandwidth estimation techniques are doomed to fail, as we show in~\ref{sec:evaluation}.

\myitem{Observation 2: Multi-threading and batching of packets introduce noise in the dispersion values.} We have observed that multi-threading at an \NF and/or batching at the receiver can cause spikes in the dispersion values as shown in Figure~\ref{fig:batch-spikes}. 
While, in principle, batching could be detected and corrected after the fact, through the periodicity of the dispersion spikes,  multi-threading further confuses the signal by causing irregular spikes depending on the distribution of load across threads. 

A straightforward approach for the estimation function $F$ given a set of dispersion values $D =\{ \delta_1..\delta_{\trainsize-1}\}$ would be to choose the minimum dispersion value. 
In this case, $F(D) =min (\ithdisp{1},\ithdisp{2},...,\ithdisp{\trainsize})$ and the capacity \cp is given by $\cp = 1/\min (\ithdisp{1},\ithdisp{2},...,\ithdisp{\trainsize})$.
Unfortunately, this estimate could lead to inaccurate results. The minimum dispersion could be affected by batching, multi-threading, or even self-induced congestion if two packets wait in the same queue, effectively modifying their dispersion. 
An alternative approach is to calculate the mean or median dispersion value across the last $N$ packets in the train since these packets are more likely to have observed the maximum frequency if the train itself has triggered it.
In this case, $F(D) =mean (\ithdisp{\trainsize-N},\ithdisp{\trainsize-N+1},...,\ithdisp{\trainsize})$ and \cp will be $\cp = 1/mean (\ithdisp{\trainsize-N},\ithdisp{\trainsize-N+1},...,\ithdisp{\trainsize})$. 
The number of packets, $N$, can be selected by finding the last step in the dispersion values using a step detection algorithm. This approach implicitly assumes that CPU frequency transitions from low to high and then remains there until the load subsides. However, we observed that in certain cases, frequency may transition back to a low frequency, as shown in Figure~\ref{fig:fs2-ex}.

\myitem{Our approach: dispersion-based \sys}
To handle the challenges mentioned in the aforementioned section, \sys first sends longer packet trains to trigger DVFS.
 Then it performs step detection~\cite{truong2020selective} on the sequence of dispersion values $D=\ithdisp{1},\ithdisp{2},...,\ithdisp{\trainsize}$ to detect the frequency transitions in the dispersion values. Second, it divides D into a set of segments $S$ at the detected steps. Third, for each segment $\disp_{i\rightarrow{j}}\in S$ from \ithdisp{i} to \ithdisp{j}, it computes the mean dispersion value of the segment.  We choose mean dispersion because it generalizes to the case of batching and multi-threading. After that, \sys picks the minimum mean dispersion values as $\disp^*$. The minimum mean dispersion will correspond to the segment of dispersion values when the frequency is maximum. Given segments $\disp_{i \rightarrow j} \in S$, we calculate $\disp^*$ as: 
\begin{align}
    \delta^* = \min_{\ithdisp{i\rightarrow j} \in S} \dfrac{\sum_{n=i}^j(\ithdisp{n}) }{j-i+1}
\end{align}
For step detection, we consider algorithms that do not require the number of steps as input since the \NF may transition to multiple frequencies, which will vary across deployments. Hence, we consider a different class of step detection algorithms~\cite{truong2020selective}, which uses penalty functions to separate noise from steps. Among these algorithms, we choose \stepDetectionAlgo~\cite{bai1997estimating} as it is widely used~\cite{chen2012parametric,fryzlewicz2014wild}.
We use the Gaussian Kernel~\cite{garreau2018consistent} as our cost function to measure variations in the data as it empirically works well. We use linear penalty~\cite{truong2020selective} for step detection based on our empirical observations. 






\subsection{Extending \sys to one-sided Control}
\label{sec:onesided}

\sys, as we have described it so far, analyzes the dispersion of packets at the attacker-controlled to infer the NF capacity.  In this subsection, we explain how \sys can be extended to a one-sided threat model as well. 

\myitem{Using routers on-path as receivers}
\name exploits the \emph{ICMP time exceeded} error response message sent by routers to an IP packet's sender if the packet has no time to reach its destination in order to learn packet dispersion without controlling the receiver. Concretely, each IP packet carries a Time-To-Live (TTL) value in its header. As routers in the Internet path forward each IP packet, they decrement this value by one. If the TTL reaches zero (expires) before the packet reaches its destination, an \emph{ICMP time exceeded} message is sent by the last router that decremented it to the sender.
The attacker: \emph{(i)} sends packets through the target NF towards a destination IP that is within the IP address space allocated to the targeted network;\footnote{IP address space allocation is public information.} \emph{(ii)} sets the TTL of the probing packets such that the TTL expires after the packet is processed by the \NF but before the packet reaches its destination; and \emph{(iii)} analyzes the dispersion of the error messages triggered the sent packets (rather than the sent packets directly). The attacker can discover the appropriate TTL by using traceroute~\cite{TracerouteLinuxMan}.\footnote{e.g., the attacker can target the last router in the path that replies to traceroute.} Traceroute is a standard measurement tool that relies on the TTL expiration mechanism and allows the sender to observe the path (the series of routers) that a packet takes to a destination. Importantly, TTL is part of a standard IP header and hence, it generalizes to various packet types.
While one could, in principle, disable ICMP time-exceeded messages to protect from \sys, that would deprive operators of a very useful debugging tool and would not necessarily protect the router's network as \sys can use any router on the path between the NF and a host.
Importantly, this technique works for any IP packet, unlike previous attempts \eg from Saroiu et al.~\cite{saroiu2002sprobe} send TCP SYN requests aiming at triggering a TCP RST from a remote host.  An alternative approach from Carter et al.~\cite{carter1996measuring} sends an ICMP ECHO REQUEST aiming at triggering an ICMP REPLY from a remote host in the target network. In both cases, the solution relies on the use of particular packet types that trigger protocol-specific behavior that might not trigger any processing from the \NF. 

\begin{figure}[t!]
    \centering
    \includegraphics[width=0.9\columnwidth]{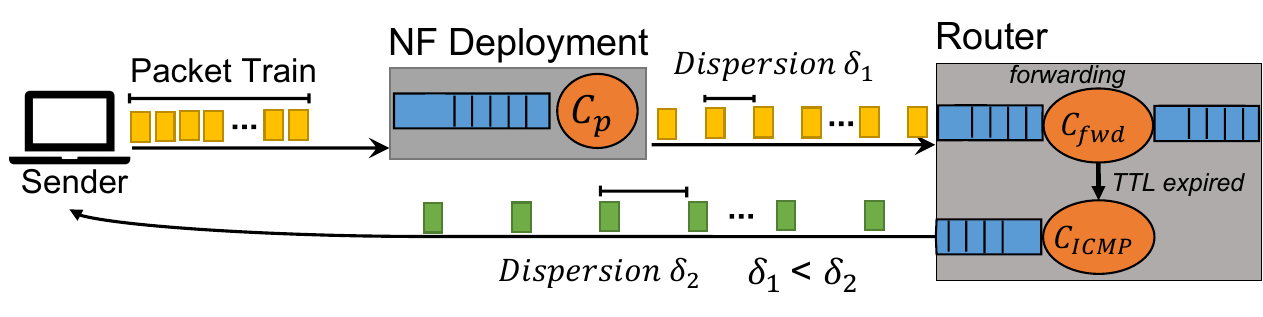}
    \caption{\small 
    One-sided threat model: While a router can be used to echo the packets back to the receiver (triggering ICMP TTL time exceeded), its processing rate may change the spacing across packets (dispersion), effectively ruining the signature.
    }
    \Description[One-sided threat model]{A router can be used to echo the packets back to the receiver (triggering ICMP TTL time exceeded). Yet, its processing rate may change the spacing across packets (dispersion), effectively ruining the signature.}
    \label{fig:router-processing}
    \vspace{-2pt}
\end{figure}

\myitem{Challenges on relying on routers for packet dispersion}
While promising, relying on \emph{ICMP time exceeded} to calculate \NF dispersion is challenging. In fact, the use of ICMP replies has well-studied limitations in the context of  traceroute~\cite{detal2013revealing,gunes2008resolving}.
We focus on three factors that are particularly impactful in our context, as they obfuscate the dispersion signals. These are related to the router processing time, contention at the router, or potential rate-limiting of ICMP traffic:

\begin{packedenumerate}

\item{\em  The router processing time might overwrite the \NF dispersion.}
In this case, the observed dispersion would correspond to the router's processing time \ie not the \NF's processing time.
We illustrate this issue in Figure~\ref{fig:router-processing}.
This is not a problem in the case of two-sided, in which routers only forward the probing packets (instead of dropping them and generating a new packet).

\item{\em The router in which the TTL expires might be congested with regular traffic, adding random delays to ICMP replies}. 
In such cases, routers prioritize regular packet forwarding from control tasks, which are often best effort. The random delays added to the ICMP packets will make their dispersion useless for an attacker attempting \NFCR.

\item{\em ICMP rate limiting may cause the routers to not respond at all.}
Finally, an operator might rate-limit the ICMP traffic~\cite{ICMPRatelimiting} for performance or security reasons. For example, in our experiments, we tried three routers in the university network and found that two of them rate-limited packets at five packets per second, as also observed by Ravaioli~\etal~\cite{ravaioli2015characterizing} 
who found that 60\% of the routers in the Internet implemented ICMP TTL exceeded rate limiting.

\end{packedenumerate}



\myitem{Our approach: one-sided \sys}
To effectively address the aforementioned limitations, we use two insights. First, while \sys needs to send large trains to trigger frequency scaling, \sys only needs the dispersion from a small number of packets to estimate \cp.
Second, to measure dispersion, \sys does not need the arrival times of consecutive packets, instead \sys can calculate the mean dispersion $\Bar{\disp}$ between two packets $p_i$ and $p_j$, which arrive at the receiver at $t_i$ and $t_j$ by $\Bar{\disp} = (t_j - t_i) / (j-i)$

Next, we explain how we use these insights to mitigate the effects of rate-limiting, router processing times, and router load.
Consider that the attacker sends a packet train of length \trainsize at time $\tau^{send}$. The attacker sets the TTL of packet $p_i$ and some other non-consecutive packet $p_j$ to expire at some router $r$. The packets arrive at the \NF, which takes time $t_p$ to process each packet. The first packet $p_1$ sees no queuing delay at the \NF. The second packet is queued for time $t_p$ and packet $p_i$ is queued for time $(i-1)t_p$ because $i-1$ packets will be processed before the packet $p_i$. Then, the packet $p_i$ arrives at the router, which will see that $p_i$ has expired TTL, and it will generate an ICMP TTL time exceeded with processing delay $t_r$. Then, let $\tau^r_i$ be the time the reply for packet $p_i$ arrives at the sender, and  $\tau^r_j$ be the time the reply for packet $p_j$ arrives at the sender. 

If $d_i$ is the total delay when packet $p_i$ left the sender and its ICMP Reply arrived at the sender, then $\tau^r_i$ is:
\begin{align}
    \tau^r_i =& \tau^{send} + d_i \label{eq:di}\\
     \text{where }d_i =& t_p + t_r + (i-1)t_p + c \nonumber
\end{align}
$c$ is for the constant transmission and propagation delay; $t_p$ is the processing delay experienced by the packet $p_i$ at the \NF; $t_r$ is the router processing delay in generating an ICMP reply; $(i-1)t_p$ is the queuing delay experienced by the packet.
We choose packet $p_j$ such that when it arrives at the router, packet $p_i$ has already been processed. This ensures that packet $p_j$ does not see any queuing at the router. Then for packet $p_j$, $\tau^r_j$ is given by:
\begin{align}
    \tau^r_j =& \tau^{send} + d_j  \label{eq:dj}\\
    \text{where } d_j = & t_p + t_r + (j-1)t_p + c \nonumber
\end{align}
At the sender, we can calculate the dispersion as follows:
\begin{align}
    \disp =  \tau^r_j -  \tau^r_i \label{eq:od}
\end{align}
Substituting Equation~\ref{eq:di} and ~\ref{eq:dj} in Equation~\ref{eq:od}, we get:
\begin{align}
    \disp =&\tau^{send} + t_p + t_r + (j-1)t_p + c  \nonumber \\ & - \tau^{send} - t_p - t_r - (i-1)t_p - c \nonumber \\
    \disp = &(j-i)t_p \label{eq:fod}
\end{align}
From Eq.~\ref{eq:fod}, the processing capacity \cp can be estimated as:
\begin{align}
    \cp = 1/t_p = (j-i)/\disp
\end{align}
%
Having explained why we do not need to trigger an ICMP error in every packet of our probe, the next natural question is how to choose these packets. To decide that, we need to look at our constraints. We want to space the packets such that when packet $p_j$ arrives at the router, packet $p_i$ has already been processed. This requires an estimation of an \emph{upper-bound} of router processing time. We can estimate processing time by sending packet pairs with the correct TTL. 
The spacing of the replies will give the router processing time. For example, in our experiments, we found that the router took 55us to process and generate ICMP replies. 

Once we have an estimation of the router processing time, we can choose the packets such that when they arrive at the router, they are at least $t_r$ apart. For example, for a router processing time of 100us, we can set the TTL after every 100 packets, assuming our send rate is 1Mpps. For a router processing time of $t_r$, and the maximum capacity $C_{max}$ that the attacker can send, we set the TTL after a gap of $g_{TTL}$ packets where $g_{TTL}$ is chosen such that $g_{TTL} > t_r \times C_{max}$, $t_r \times C_{max}$ gives the number of packets arriving at the router in time $t_r$. 
Note that the \NF processes packets before they arrive at the router. Thus, the actual duration between the packets will be much greater than $t_r$. 
For example, consider an \NF, which processes each packet in 5us, and a router takes 100us to generate an ICMP reply. If the sender sets the TTL of every 100th packet, then if the first packet arrives at the router at time t, the 101st packet will arrive at $((5us) + (100)(5us))\times t$, which is much greater than 100us. 
Importantly, the estimation of the processing in the router need not be extremely accurate. In fact, we only need to make sure that the packets are spaced out more than the processing time.

%


\hspace*{-\parindent}%
\begin{figure*}
\begin{minipage}[t!]{\textwidth}
\begin{minipage}{0.73\textwidth}
       \begin{subfigure}[b]{0.55\columnwidth}
\includegraphics[width=\columnwidth]{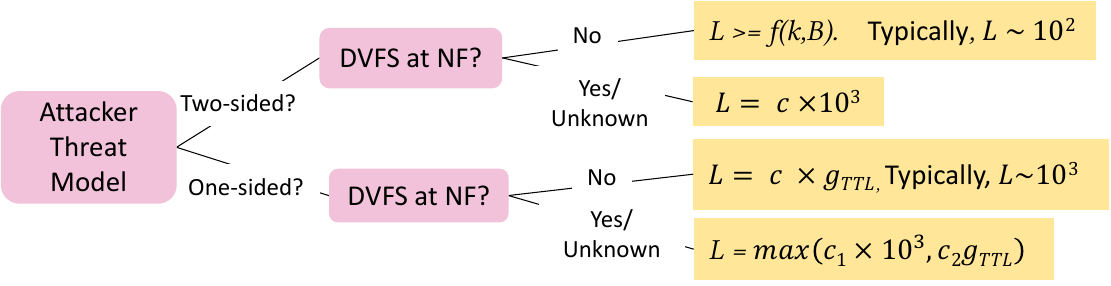}
     \caption{ \label{fig:decision-tree}}
   \end{subfigure}
   \hfill
    \begin{subfigure}[b]{0.4\columnwidth}
      \includegraphics[width=\columnwidth]{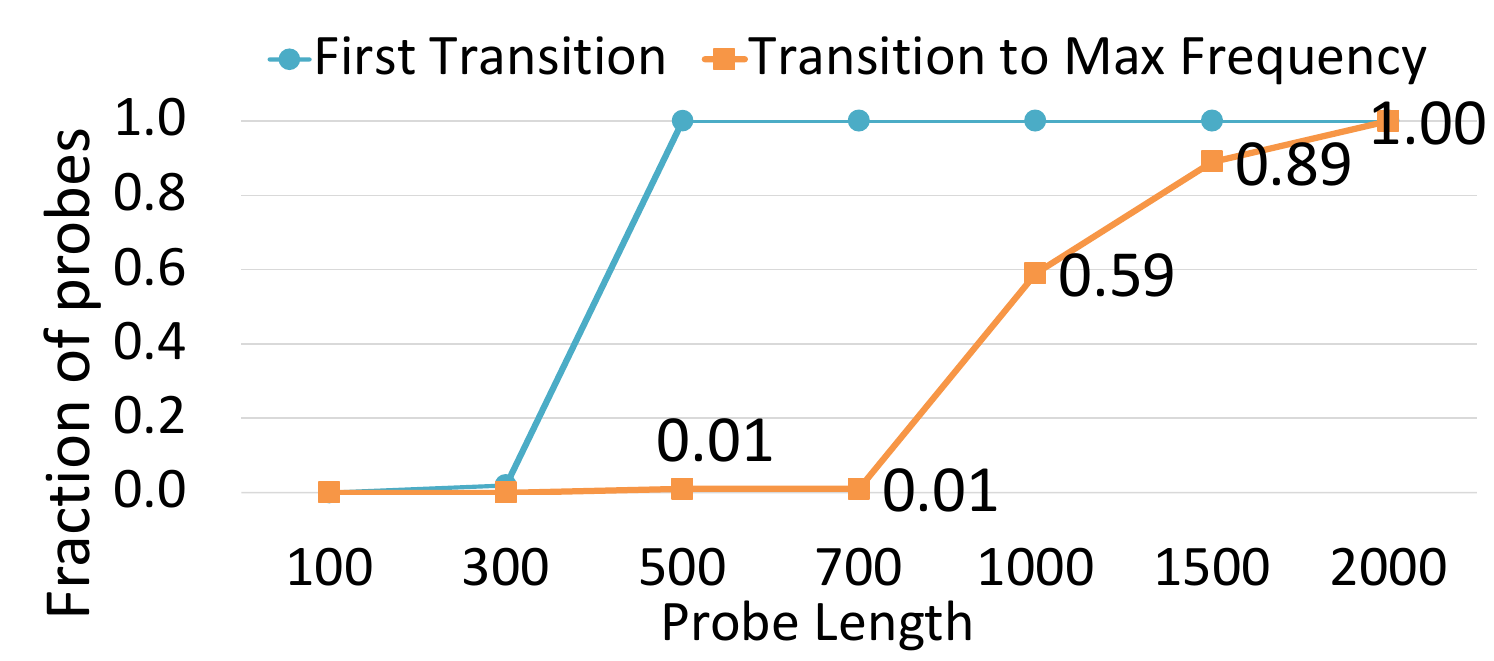}
    \caption{\label{fig:fs-length}}
    \end{subfigure}
    \caption{(a) \NFCR is challenging because of factors such as whether the attacker controls nodes on both sides of the \NF, whether DVFS is enabled in the \NF deployment, packet batching \etc Here, $\trainsize$ is the probe length,  $k$ is the number of threads, $B$ is the batch interval, $g_{TTL}$ is the packet gap needed for the one-sided case, and $c,c_1,c_2$ are constants. (b) Poorly-selected train size may not trigger a transition to the maximum frequency; e.g., a train of 1000 packets has only 59\% chances of observing the \emph{maximum} frequency.
      }
    \Description[]{}
\end{minipage}
  \hfill
\begin{minipage}{0.25\textwidth}
    \centering
    \includegraphics[width=\columnwidth]{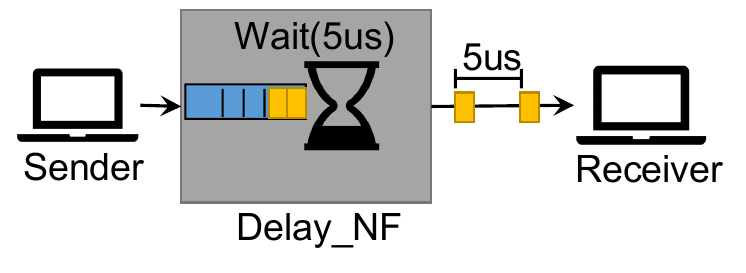}
    \captionof{figure}{\small  \label{fig:dnf-setup}Our testbed comprises three servers acting as the sender, the receiver and the \NF. The \dnf delays each packet by a predefined amount. For example, the \dnf here delays each packet by 5 $\mu$s.}
  \end{minipage}
\end{minipage}
\end{figure*}

\subsection{End-to-end View}\label{sec:e2e}

In this subsection, we discuss how we can configure \sys 
 depending on various deployment settings and constraints. The key configuration parameter here is the ideal probe length $L$ for the attacker, \ie the minimum number of packets for a good estimation. 
 This, in turn, depends on all the factors we have mentioned so far, including \NF's deployment and whether the attacker controls the receiver and hardware configuration.
  Figure~\ref{fig:decision-tree} summarizes our insights into the various factors that play a role in this decision. 

Starting from the bottom branch in Figure~\ref{fig:decision-tree},
if the attacker is one-sided, meaning they only control the senders and rely on a router for receiving some form of a reply, she would need to send $c_1  \times g_{TTL}$ where  $g_{TTL}$ is the gap between packets for which the attacker sets the TTL, and $c_1$ a constant which determines the number of TTL Exceeded packets the attacker will receive.  
Remember that in the one-sided case, to deal with routers with longer ICMP router processing times and with ICMP rate-limiting, \sys spaces out the packets with the correct TTL. This gap determines the length of the probe. In our experiments, $30\times g_{TTL}$, with $g_{TTL}$ of 100 packets (total of 3000 packets) yields accurate capacity estimations.  In the presence of DVFS, the attacker will again need to send packets in the order of thousands to first trigger DVFS to see the maximum capacity, as shown in Figure~\ref{fig:fs-length}.

If the attacker is two-sided and the \NF does not use DVFS (topmost branch), the probe length will depend on the number of threads in the \NF or/and the batch interval in case of packet batching. First, to capture the effect of multi-threading, the attacker will need to send packets $c \times k$ packets at the \NF with $k$ threads, where $c$ is some constant. Intuitively, the value of $c$ is proportional to the number of threads. 
Second, to handle the noise introduced by batching, attacker will need to send packets such that the total arrival time of packets at the receiver is larger than the batch interval. 
Batching is usually done in microseconds with values typically around 100us~\cite{liang_2017}. Hence, in this branch of the decision tree, a few hundred packets will give a good estimate, assuming that the number of threads is also in the hundreds or less.

If the attacker is two-sided and the \NF uses DVFS, the packets needed to trigger DVFS will be in thousands. Similarly, in the one-sided case, the number of packets typically needed would be in the thousands. In these cases, the number of threads and batching will not play a role in determining the probe length.  

Given this context, we distinguish two representative \sys configurations based on the decision tree in Figure~\ref{fig:decision-tree}. The two configurations differ in effectiveness and stealthiness. \\
\myitem{\syslarge} uses \dvfslength packets to probe an \NF, it will accurately estimate the capacity of \NF with or without optimized deployments.
Importantly, \syslarge is also effective with the one-sided threat model. As we show in Section~\ref{sec:evaluation}, \syslarge  can accurately estimate the capacity of diverse \NF deployments within \sysonesidederror error in the Internet in a two-sided and one-sided threat model. Moreover, \syslarge outperforms link-bandwidth estimation baselines by \syslargedelta.
We do not claim that \syslarge will be enough to detect accurately in all cases.

\myitem{\syslite} uses 100 packets to probe an \NF. It will work for simple \NF deployments, where DVFS is disabled. Due to its tiny network footprint \syslite is very stealthy. 
While \syslite is less accurate than \syslarge, it is still powerful enough to estimate the capacity of a commercial NF deployed in AWS.
\syslite falls in the topmost branch of the decision tree.
Since \syslarge is strictly more accurate, we expect an attacker to use \syslarge, unless they have a strict budget or side-channel information about the NF deployment. 
We thoroughly evaluate both attacks in Section~\ref{sec:evaluation}.
\section{Measurement Framework}
\label{sec:measurements}

For our probing and estimation technique to work, we need a robust measurement infrastructure to get accurate, fine-grained, and noise-free dispersion measurements. To achieve this, we require:\\
\textbf{Fast sending rate:} The sending rate should be faster than the \NF capacity \cp to ensure that packets are queued when they arrive at the \NF.\\ 
\textbf{Accurate timestamping:} Packets should be timestamped at microsecond precision when they arrive at the receiver to capture the dispersion accurately. 

We build a test setup (Figure~\ref{fig:dnf-setup}) to systematically verify our sender and receiver setup. We create a dummy \NF, namely \dnf, with a configurable processing time, implemented as a sleep operation for a configurable time window upon receiving a UDP packet 
We implemented \dnf using Click ~\cite{kohler2000click} and it serves as the ground truth for dispersion. For each experiment, we send packet trains of 10 packets and report the error in the mean dispersion. For packet generation, we evaluate tools such as Hping3, TCPReplay, and MoonGen. For timestamping at the receiver, we test TCPDump and DPDKcap. We also explore certain sender (\eg DVFS~\cite{DVFS}) and receiver side configurations (\eg RSS~\cite{ReceiveSideScaling}, interrupt throttling) that may affect the dispersion. Next, we present our findings, which guide \sys design.

\begin{figure}
    \centering
   \begin{subfigure}[b]{0.48\columnwidth}
    \includegraphics[width=\columnwidth]{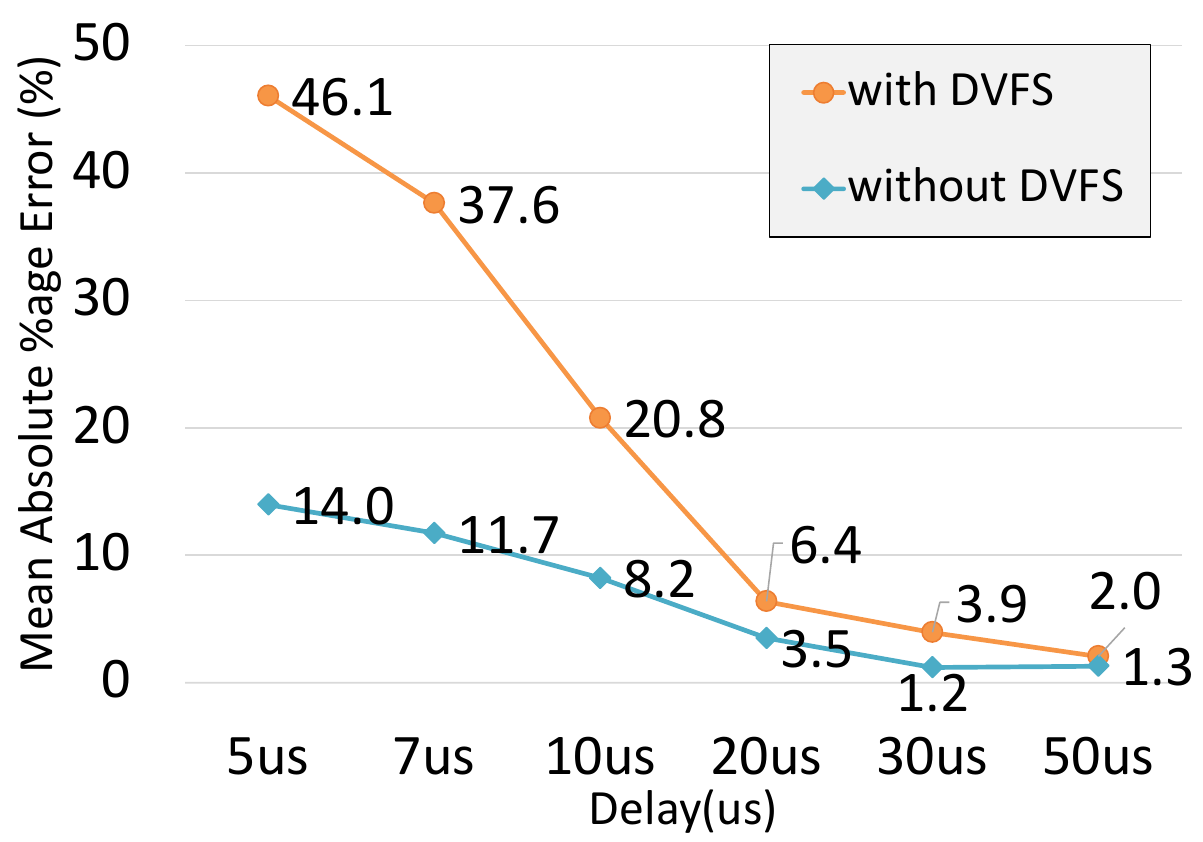}
     \caption{ \label{fig:fs-sending}}
   \end{subfigure}
    \begin{subfigure}[b]{0.48\columnwidth}
      \includegraphics[width=\columnwidth]{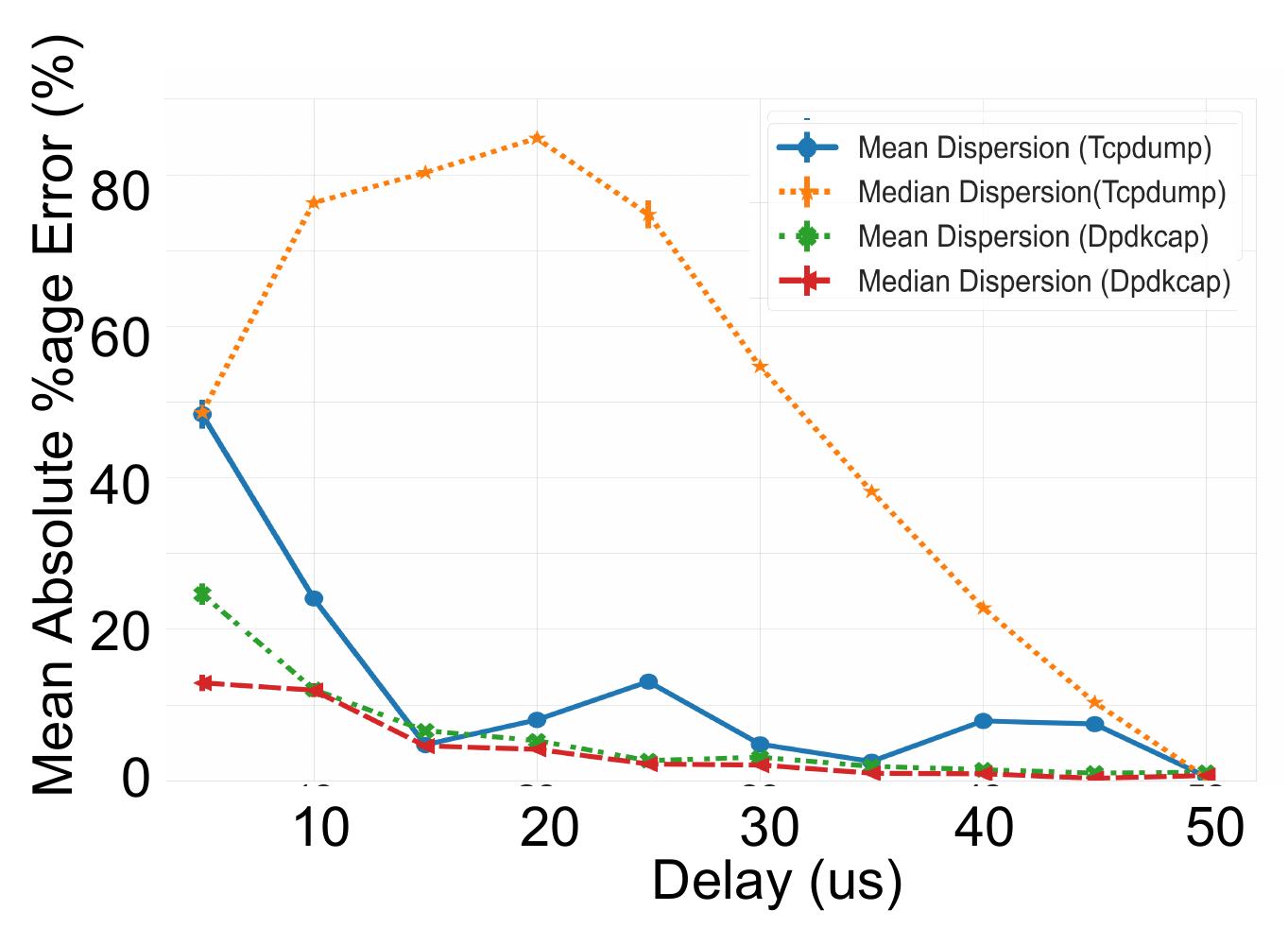}
      \caption{\label{fig:mape-disp}}
    \end{subfigure}
    \caption{\small
    (a) Frequency scaling (DVFS) at the sender results in high error for lower processing times. Disabling frequency scaling at the sender significantly reduces this error.   (b)TCPDump results in high error in the measured dispersion because of interrupt throttling.
    }
    \label{fig:my_label}
    \Description[Insight on Measurement Framework]{(a) Frequency scaling (DVFS) at the sender results in high error for lower processing times. Disabling frequency scaling at the sender significantly reduces this error.   (b)TCPDump results in high error in the measured dispersion because of interrupt throttling.}
    \vspace{-5pt}
\end{figure}

\myitem{\sys turns off DVFS at the sender to send packets at a faster rate}
Frequency scaling at the sender node can affect the send rate, besides the NF, as we explained in~\ref{sec:dispersion}.
Figure~\ref{fig:fs-sending} shows the error in measured dispersion for different delay values of the \dnf.  At low CPU frequency, the sender is not able to send packets as fast. Hence, in the presence of frequency scaling at the sender, the error in measured dispersion can reach up to 46.1\% for smaller delay values.  
Note that we observed a similar trend for kernel-based tools and MoonGen. 
Hence, we turn frequency scaling off at the sender node. All bandwidth estimation tools do not consider this factor and hence may get affected, as we show later in the evaluation section~\ref{sec:evaluation}.

\myitem{\sys uses DPDKcap to avoid the negative effects of interrupt throttling on the timestamping of received packets.  }
On the receiver side, using DPDKcap reduces the noise in the measured dispersion and has up to 2X lower error compared to  TCPDump, as shown in Figure~\ref{fig:mape-disp}. 
Essentially, TCPDump entails high errors because of interrupt throttling and delays in the kernel networking stack.  While prior work~\cite{prasad2004effects,yin2016can} tries to handle the effects of batching by estimating the batching interval and correcting for it, this is hard in the presence of adaptive interrupt coalescence~\cite{IC}. DPDKcap, on the other hand, bypasses the kernel and the need for interrupts via userspace polling. 

\section{Evaluation}
\label{sec:evaluation}
We evaluate the accuracy and the stealthiness of \sys in two representative configurations, namely \syslarge and  \syslite.

We find that \emph{(i)} \syslite and \syslarge are up to \syslitedelta and \syslargedelta more accurate in controlled experiments compared to the equivalent baselines sending the same number of packets, and are more stealthy than the baselines (\sys is rarely detected by the ruleset of known commercial firewalls); \emph{(ii)}\syslite is stealthy yet accurate only on a subset of the NF deployments compared to \syslarge but still able to accurately estimate the capacity of a commercial \NF in AWS  within \sysliteaws of error; \emph{(iii)} \sys's step-detection improves its accuracy by 25x compared to other estimation baselines while its measurememt improves \sys's performance by up to 33x; and \emph{(iv)} \sys also estimates accurately with up to 10\% error.

We explain our evaluation in detail below, starting with our methodology.
\begin{figure}[t!]
    \centering
   \centering
    \includegraphics[width=0.85\columnwidth]{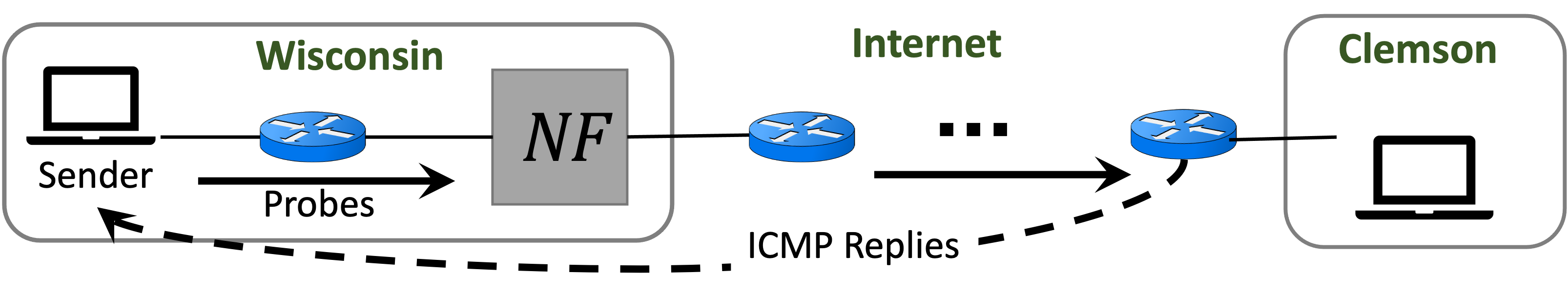}
\caption{\label{fig:internet-setup} Our Internet setup forwards packets through the Internet after they have been processed by the \NF. This is the worst-case scenario for the attacker (and \sys) as the Internet noise can affect the dispersion signature of \NF. In the one-sided experiments, we send probes to the Clemson node but they expire on path, causing ICMP replies to the sender (shown in dotted).}
\Description[Internet Setup]{Our Internet setup forwards packets through the Internet after they have been processed by the \NF. This is the worst-case scenario for the attacker (and \sys) as the Internet noise can affect the dispersion signature of \NF. In the one-sided experiments, we send probes to the Clemson node but they expire on path, causing ICMP replies to the sender(shown in dotted).}
\end{figure}

\subsection{Methodology}
\myitem{Probing baseline:} 
We evaluate our approach against a popular and widely-used  rate-based link bandwidth-estimation technique, namely \slops in which the attacker uses binary search for \NFCR~\cite{jain2002end,kachan2015available} between 0-500Kpps. 
\slops sends packets to the NF at an initial probing rate (250Kpps) and uses one-way delay of received packets to see if the probe
rate (pps) caused congestion, indicating that the rate was greater than the NF capacity. If so, it
adjusts the probe rate accordingly using a binary search
approach.  \slops keeps probing until the difference
between the minimum and maximum rate is 1000pps. To
find an increasing trend in the relative one-way delay values for  \slops, we
use the technique implemented by Pathload~\cite{jain2002end}. 
Moreover, we use the same probe length for the baseline and for \sys.

\myitem{Estimation baselines:}
To evaluate the estimation technique used by \sys, we compare with two baseline approaches adopted from the link bandwidth estimation literature \emph{(i)} 
\myitem{\bone} which takes the mean dispersion of the entire train~\cite{dovrolis2001packet}; and \emph{(ii)} \myitem{\btwo} which takes the median dispersion~\cite{li2008wbest}. 

\myitem{Network functions:} We use five \NFs.
Both \snortrl and \surrl are realistic TCP SYN rate-limiting \NFs which we implemented in Suricata~\cite{Suricata} and Snort~\cite{SnortNetworkIntrusion} respectively.
Both \NFs track the number of SYN packets per flow and drop the SYNs of flow that have exceeded a threshold. For \snortrl and \surrl, we send SYN packets.
\surbl is a deny list containing rules provided by the emerging threats database~\cite{ProofpointEmergingThreats}. We only send UDP traffic to \surbl to trigger the rules concerning UDP traffic. We implemented \surbl in Suricata~\cite{Suricata}. 
\surblmulti and \surrlmulti are multi-threaded versions of \surbl and \surrl respectively.

\begin{table}[!t]
\small
    \centering
    \begin{tabularx}{\columnwidth}{p{1.5cm}llll}
    NF  ($\mu s$) & Controlled &  One-sided & Internet2 & Internet\\
        \hline
      \snortrl & 67500 & 21800 & 21000 & 132388\\
      \surbl & 142600  & 143420& 134640 & 144940 \\
      \surrl & 200880 & 225140& 193750 & 227726\\
      \surblmulti & 227720  & 216000& 224680 & 246679\\
      \surrlmulti & 336060  & 309140&360460 &348795 \\
      \bottomrule
    \end{tabularx}
    \caption{The measured ground-truth processing capacity (packets per second) of our NFs in different settings}. 
    \label{tab:nf-benchmark}
\end{table}

\myitem{Ground-truth measurements:} 
We measure the ground-truth capacity for each setup and \NF separately, as it can be affected by hardware and software factors (\eg processor, NIC).
To measure the ground-truth, we send packets to a receiver node via each \NF using Scapy $\mathit{sendpfast}$ ~\cite{rohith2018scapy}. We do a binary search on the sending rate until we find the maximum rate that makes the input rate of the \NF node equal to its output rate (\ie the maximum sending rate without packet drops at the \NF). 
We repeat this exercise five times and choose the median as the ground truth.
We summarize our results in Table~\ref{tab:nf-benchmark}.
We repeat each experiment 100 times and report the Median Absolute Percentage Error (MdAPE) in the estimated processing capacity compared to the ground truth (Table~\ref{tab:nf-benchmark}). 



\subsection{Controlled experiments}

\begin{figure*}[t!]
\small
\centering
    \begin{subfigure}[b]{0.33\textwidth}
        \includegraphics[width=\columnwidth]{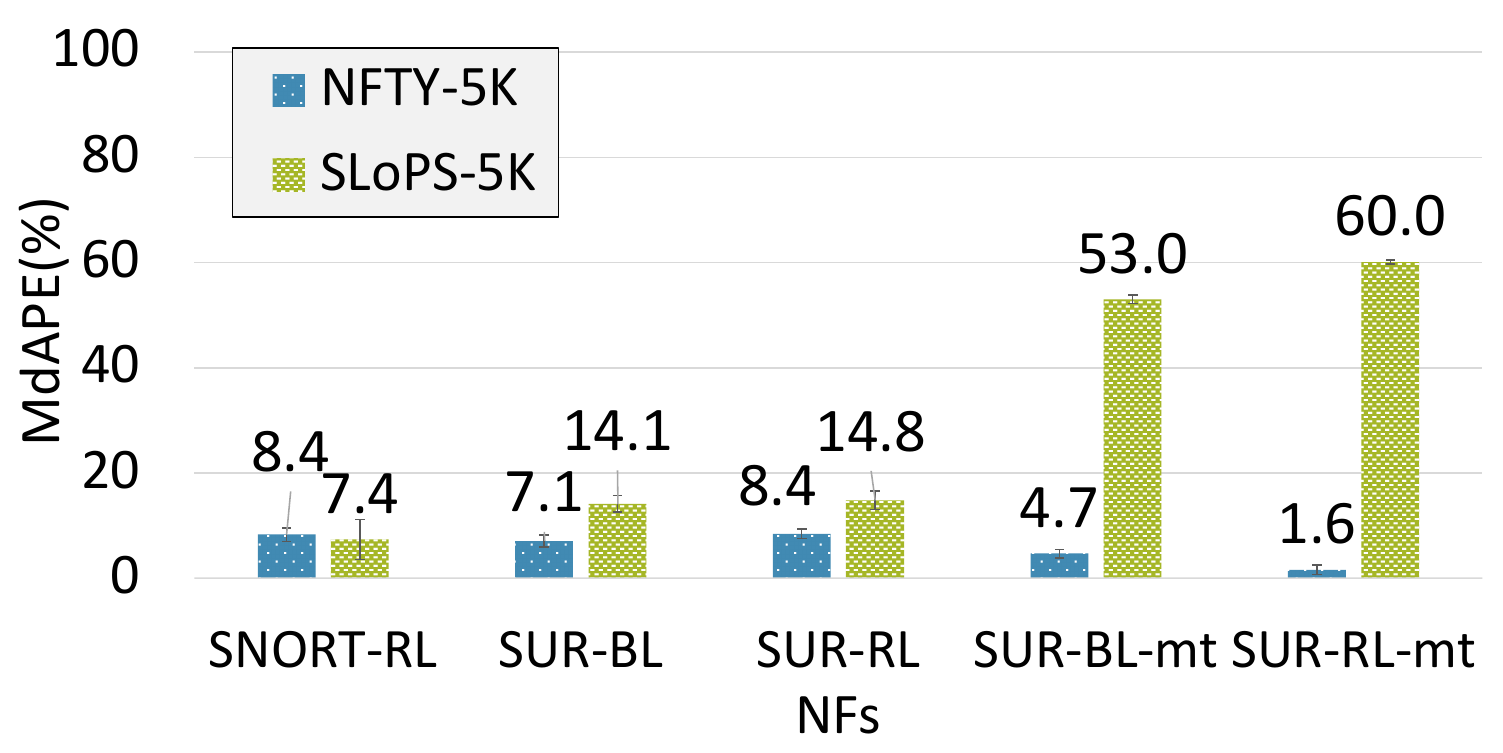}
        \caption{\label{fig:nfty-large-lab}}
    \end{subfigure}
    \begin{subfigure}[b]{0.33\textwidth}
        \includegraphics[width=\columnwidth]{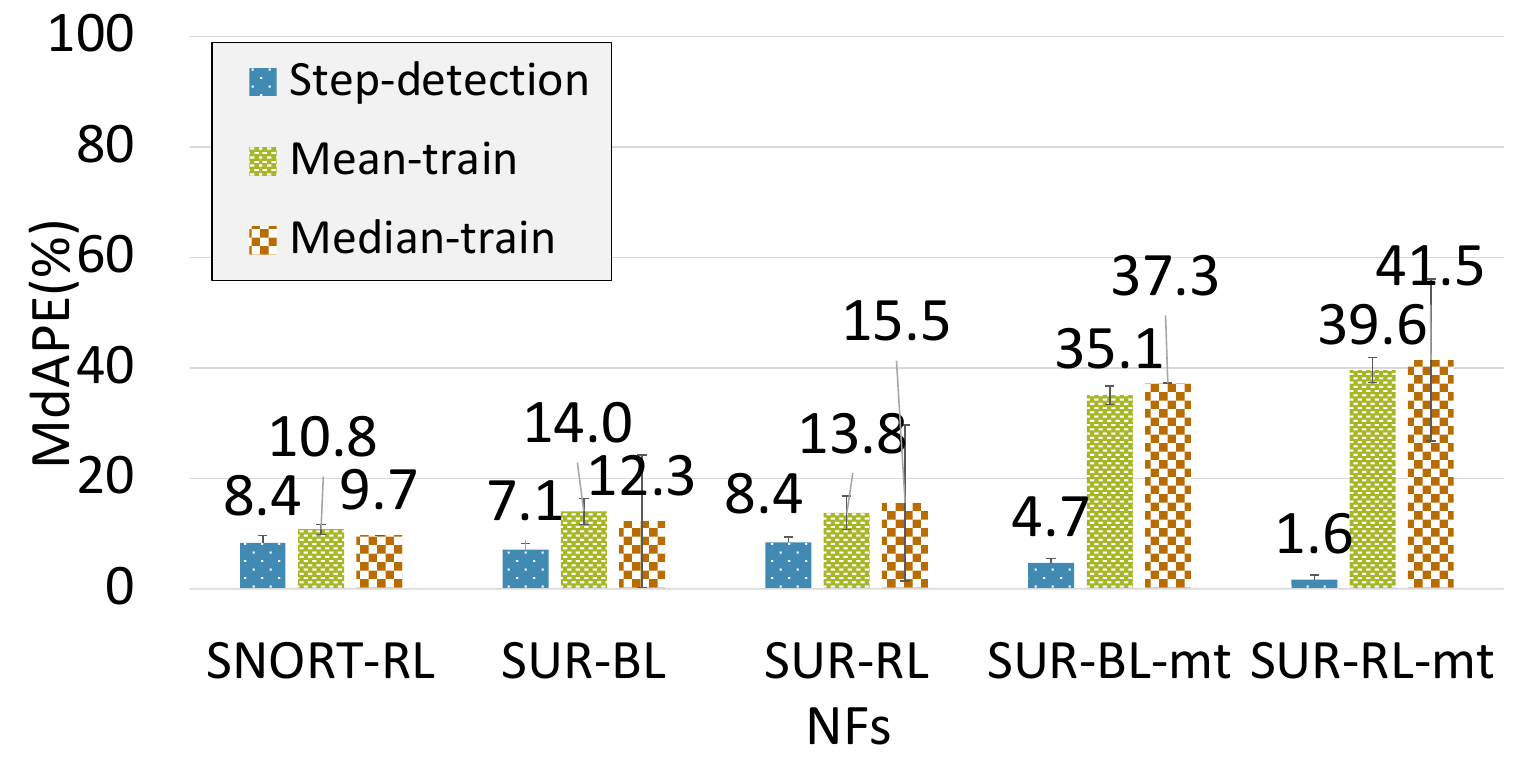}
     \caption{\label{fig:step-detection-result}}
     \end{subfigure}
      \begin{subfigure}[b]{0.32\textwidth}
       \includegraphics[width=\columnwidth]{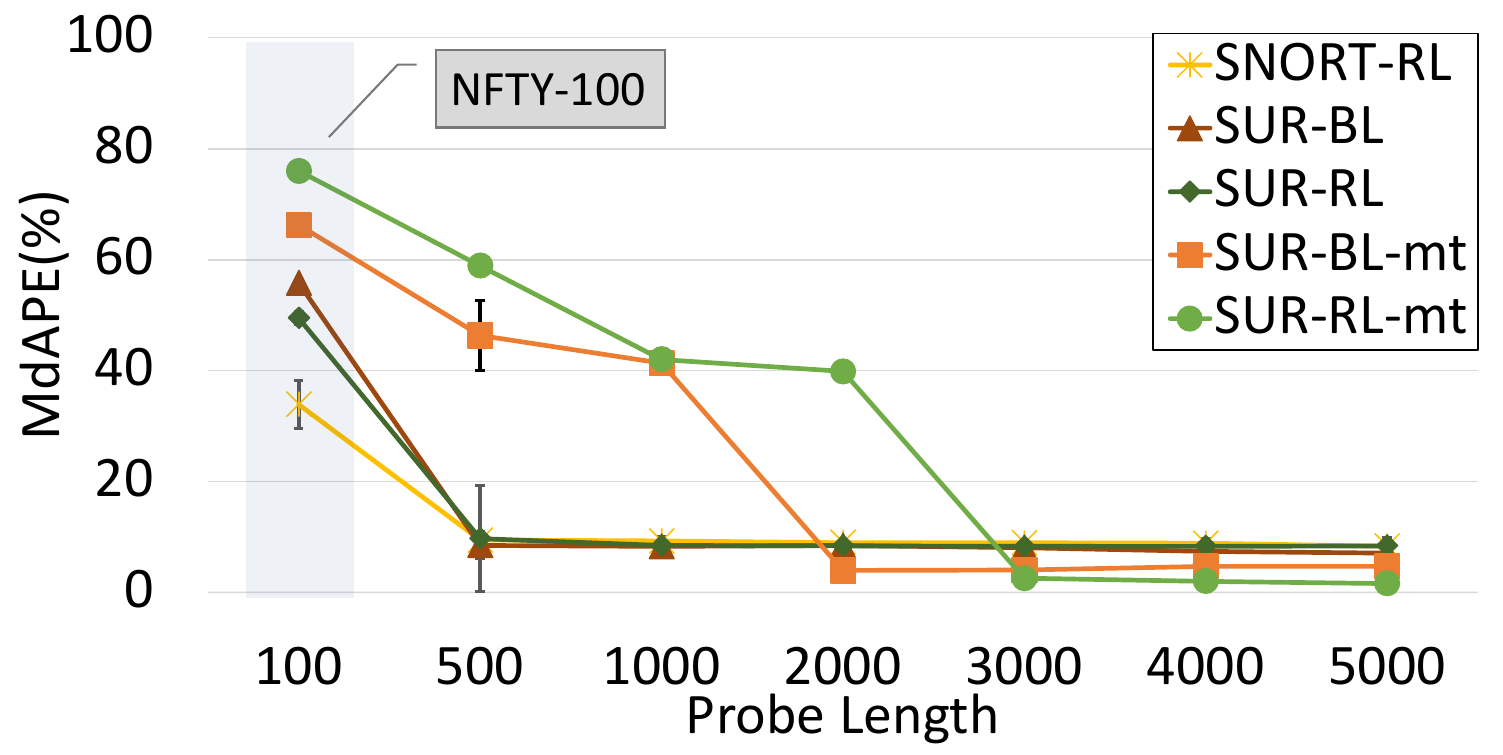}
       \caption{\label{fig:budget}}
      \end{subfigure}
      \vspace{-1pt}
    \caption{(a) \syslarge is up to \syslargedelta more accurate than the baseline in optimized \NF deployments (b) Step detection in dispersion values reduces the MdAPE of \syslarge by up to \stepdetectiondelta. (c) As we increase the probe length, the accuracy of \sys improves.}
    \Description[\syslarge Performance]{(a) \syslarge is up to \syslargedelta more accurate than the baseline in optimized \NF deployments for the same probes length.  (b) By applying step detection to the dispersion values, the MdAPE of \syslarge is reduced by up to \stepdetectiondelta.  (c) As we increase the probe length, the accuracy of \sys improves.}
\end{figure*}

We evaluate  \syslarge on optimized \NF deployments \ie with DVFS enabled, 
before we evaluate \syslite, for simplified ones. 
 We use three nodes (Intel Haswell EP nodes with two 2.60 GHz 10-core CPUs) in CloudLab~\cite{duplyakin2019design}, a sender, a receiver, and an \NF connected such that the sender communicates to the receiver via the \NF node.  

\begin{figure*}[t!]
\small
\centering
    \begin{subfigure}[b]{0.33\textwidth}
        \includegraphics[width=\columnwidth]{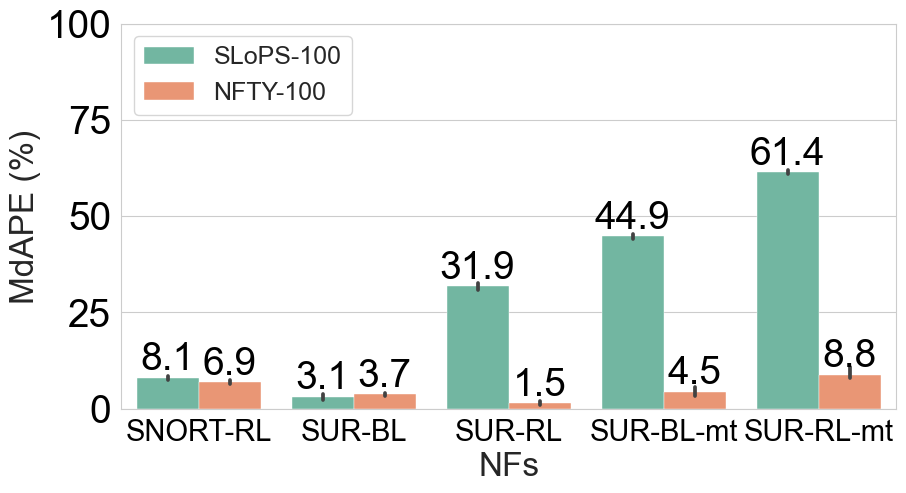}
        \caption{\label{fig:nfty-lite-result}}
    \end{subfigure}
    \begin{subfigure}[b]{0.33\textwidth}
        \includegraphics[width=\columnwidth]{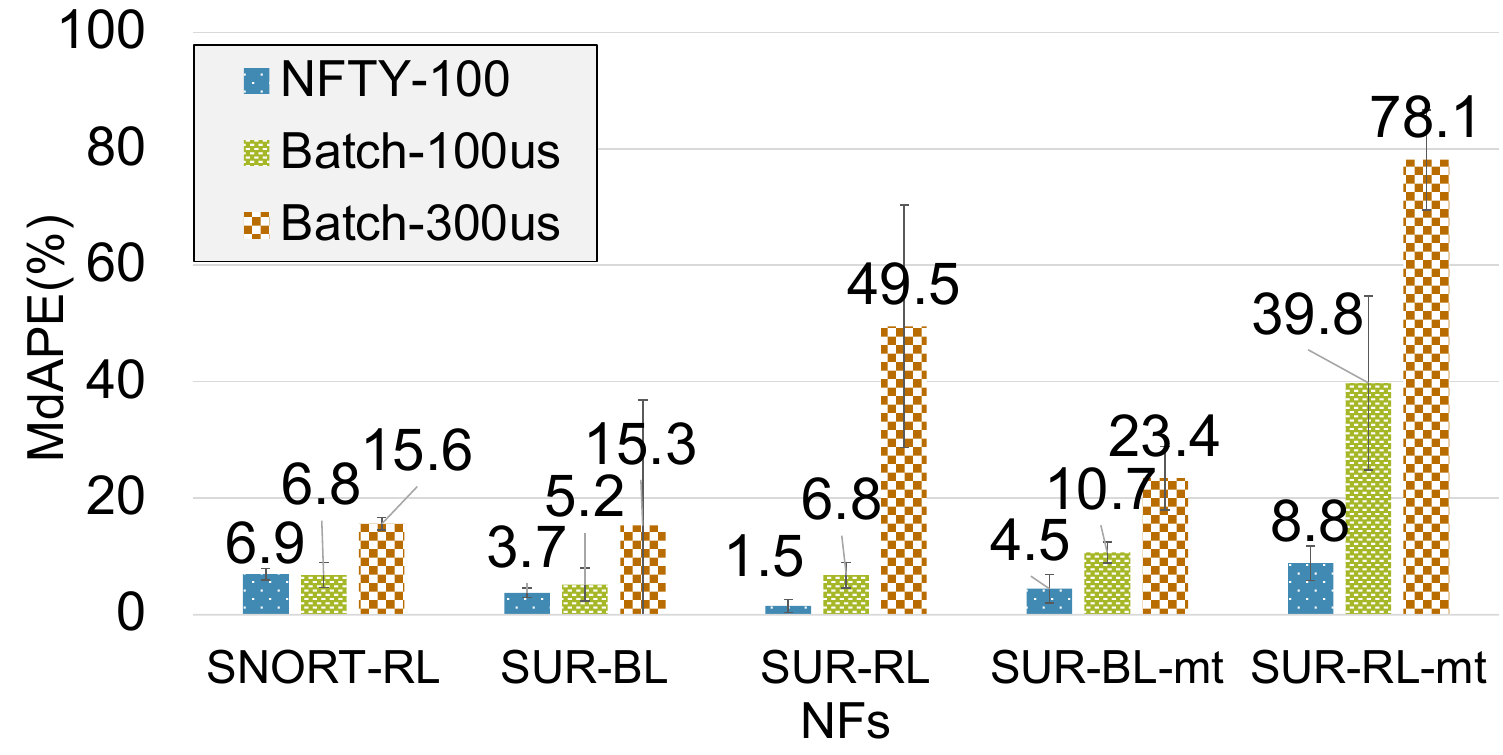}
     \caption{\label{fig:batching}}
     \end{subfigure}
      \begin{subfigure}[b]{0.32\textwidth}
       \includegraphics[width=\columnwidth]{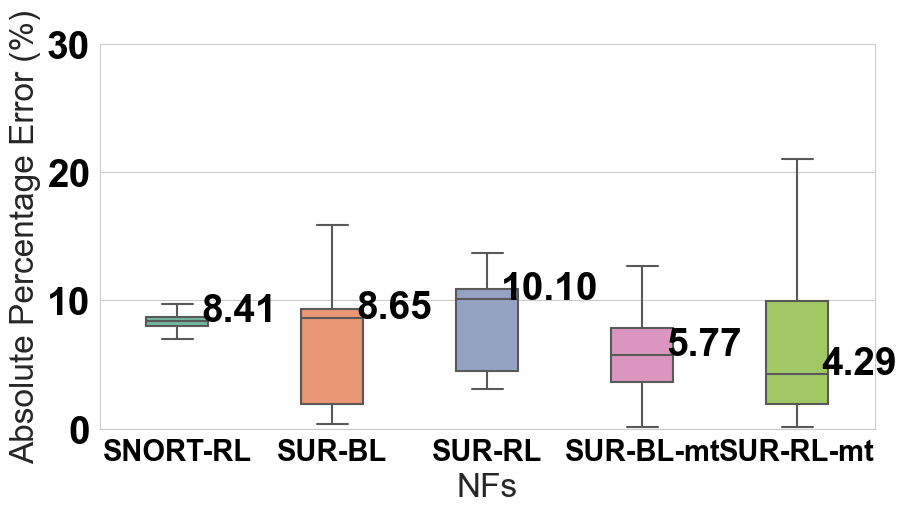}
       \caption{\label{fig:one-sided}}
      \end{subfigure}
    \caption{(a) \syslite results in up to \sysliteerrorlab error with the optimized measurement infrastructure. \slops-100 results in high error (up to 61\%) for faster \NFs as the sender cannot send packets faster than the \NF because of DVFS at sender. (b) When we use \syslite in an optimized measurement infrastructure, the MdAPE is reduced by up to \measurementoptdelta. (c) \sys results in 4\%-\sysonesidederror MdAPE with one-sided control. The spread in error is a bit high for fast \NFs compared to \snortrl, \ie  the slowest \NF. }
    \Description[\syslite Performance]{(a) \syslite results in 3\% to \sysliteerrorlab error in the estimated processing capacity with the optimized measurement infrastructure. \slops-100 results in high error (up to 61\%) for faster \NFs as the sender cannot send packets faster than the \NF because of DVFS at sender. (b) When we use \syslite in an optimized measurement infrastructure, the MdAPE is reduced by up to \measurementoptdelta.  The improvement is significant for the larger batch interval (batch-300us) as compared to the smaller interval (batch-100us). (c) \syslite's error remains below \sysliteinterneterror in the internet. The spread is a bit higher than the controlled experiments for very fast \NFs due to the noise added by the Internet.  }
\end{figure*}

\myitem{\syslarge outperforms probing baselines by 2x-\syslargedelta while sending less packets in total.}
Figure~\ref{fig:nfty-large-lab} shows the MdAPE in the estimated capacity by \syslarge results in 2\%-\syslargeerrorlab error and outperforms the baseline. 
The improvement is more prominent in the faster \NFs, compared to \snortrl, which is slower (see Table~\ref{tab:nf-benchmark}). 
The baseline (\slops) also ends up sending 9x more packets (median in NFs) than \syslarge, due to its iterative probing. The baseline's poor accuracy is due to its obliviousness to the fluctuation of one-way delays caused by the NF's optimized deployment (\ie frequency transitions). Moreover, for multi-threaded NFs (last two bars), the baseline triggers only a single thread because both the multi-threaded \NFs assign packets to threads per flow.
Importantly, both techniques use the optimized measurement infrastructure we describe in \S\ref{sec:measurements}. The baseline's accuracy further degrades with an unoptimized measurement infrastructure (as shown in Figure~\ref{fig:nfty-lite-result}). 

\myitem{Step-detection improves \syslarge accuracy by up to \stepdetectiondelta}
To investigate the benefit of step detection, we compare \syslarge with the estimation baselines, \bone and \btwo in Figure~\ref{fig:step-detection-result}. The reduction in MdAPE is significant. For  \snortrl the improvement over the baseline is less compared to the other NFs because \snortrl changes to maximum frequency sooner than the other \NFs. In effect, \bone is closer to the one corresponding to the maximum frequency. \btwo precision degrades further for multi-threaded \NFs, because of the fluctuations in queueing delays for subsequent packets it causes.

\myitem{Longer probe lengths are critical for NFs with DVFS.} 
Figure~\ref{fig:budget} shows the effect of different probe lengths on \sys's error in the presence of DVFS. Longer trains for \sys result in lower error (MdAPE) because they always trigger the transition to the maximum frequency. In contrast, small trains never trigger the transition to maximum frequency. Thus, if the attacker were to perform \NFCR with \syslite in the presence of DVFS, she would have a high error. 

\myitem{\syslite accurately estimates capacity within \sysliteerrorlab of error.}
As \syslite cannot estimate the capacity of NFs whose deployments are optimized with DVFS (Figure\ref{fig:budget}), we evaluate \syslite for simpler \NF deployments. \syslite estimates the capacity of these with 2\% to \sysliteerrorlab error as shown in Figure~\ref{fig:nfty-lite-result}. 
We compare \syslite with \slops-100. For \syslite, we turn DVFS off at sender, as with small number of packets, its effect becomes significant. For the baseline, DVFS is not off at the sender.
The error for \slops-100 increases up to ~61\%. Specifically, for \slops-100, the error is huge for \NFs (\surrl to \surrlmulti) where the attacker cannot reliably send at rates higher than the \NF processing speed. 
If an attacker were to use \slops-100 without our sender-side optimization, she would incur a huge error.

\myitem{\sys's measurement optimizations improve the accuracy of \syslite  by up to \measurementoptdelta}
In the previous experiment, we hinted that the most impactful factor in  \syslite accuracy is the measurement infrastructure (see Sec. \S\ref{sec:measurements}). To verify that, we compare the accuracy of \syslite in an optimized and unoptimized measurement infrastructure in Figure~\ref{fig:batching}. For the unoptimized measurement infrastructure, DVFS is enabled at the sender, and we show two receiver configurations, one with a batch interval of 100us and the other with 300us. For the optimized measurement infrastructure, DVFS is disabled at the sender, and there is no packet batching. 
As shown in Figure~\ref{fig:batching}, our measurement infrastructure optimizations in \syslite, reduce its MdAPE by upto \measurementoptdelta. 

\myitem{\sys is stealthier than \slops while being more accurate}
To evaluate the stealthiness of \sys, we use the ruleset of known commercial firewalls Palo Alto Networks (PANW)~\cite{panw_2024}, Juniper (JNPR)~\cite{juniper}, Fortinet (FTNT)~\cite{fortinet_2024}, and Snort community rules~\cite{snortcommunity_2024}. We implement these rulesets in Suricata, and record if these rulesets detect \sys and \slops. Table~\ref{tab:detect} summarizes our results for TCP SYN traffic processed by \surrl, \surrlmulti, and \snortrl and UDP traffic processed by \surbl, and \surblmulti. A tick indicates that the firewall was able to detect the attack. A cross indicates that the attack is not detected. Overall, \sys is less detectable than the baseline.

     
\begin{table}[t!]
\renewcommand{\arraystretch}{0}
\small
    \centering
    \begin{tabularx}{\columnwidth}{@{}p{1.315cm}p{0.6cm}p{0.5cm}p{0.5cm}p{0.5cm}|p{0.6cm}p{0.5cm}p{0.5cm}p{0.5cm}@{}}
     & PANW & JNPR & FTNT & Snort & PANW & JNPR & FTNT & Snort\\
      \midrule
    & \multicolumn{4}{c}{TCP SYN} &  \multicolumn{4}{c}{UDP}\\
     
    \syslarge & \redcross & \redcross & \greentick & \redcross  & \redcross & \greentick & \greentick & \redcross \\
    \slops-5K & \greentick & \greentick & \greentick & \redcross & \greentick & \greentick & \greentick & \redcross\\
    \syslite & \redcross & \redcross & \redcross & \redcross & \redcross & \redcross & \redcross & \redcross\\
    \slops-100 & \redcross & \redcross & \redcross & \redcross &  \redcross & \greentick & \redcross & \redcross \\
    \midrule
     
    \end{tabularx}
    \caption{\sys is  stealthier than \slops while being more accurate as shown by the \greentick  where the attack is detected by different rulesets}. 
    \label{tab:detect}
    \vspace{-5pt}
\end{table}

\subsection{One-sided experiments}
To evaluate \sys under the one-sided model, we use \syslarge.
We illustrate the setup of this experiment in Figure~\ref{fig:internet-setup}. The sender and the \NF are in the Wisconsin cluster, and they send packets to a public server in the Clemson cluster. The TTL is set such that it expires in 4 hops after being processed by the NF. We choose $g_{TTL}$ (\ie the gap in packets for which we set the correct TTL) to be 100 packets.

\myitem{\syslarge is usable under the one-sided threat model.} 
We summarize our results in Figure~\ref{fig:one-sided}. 
We see that
\syslarge's MdAPE remains around \sysonesidederror, even under the one-sided threat model. Faster \NFs such as \surrlmulti show a higher spread as in the case of two-sided Internet experiments. 

\begin{figure*}[t!]
    \centering
    \begin{subfigure}[b]{0.5\columnwidth}
  \centering
    \includegraphics[width=\columnwidth]{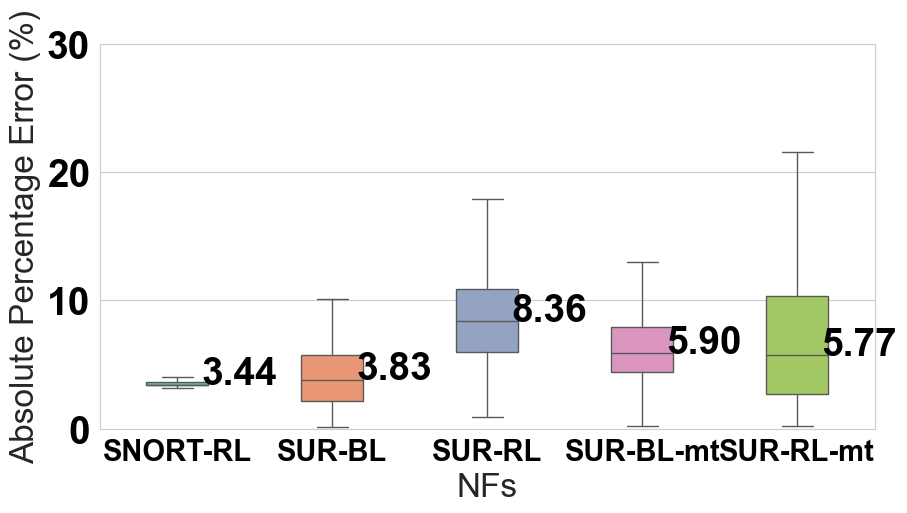}
    \caption{\label{fig:internet-fs-results}}
\end{subfigure}    
 \begin{subfigure}[b]{0.5\columnwidth}
       \includegraphics[width=\columnwidth]{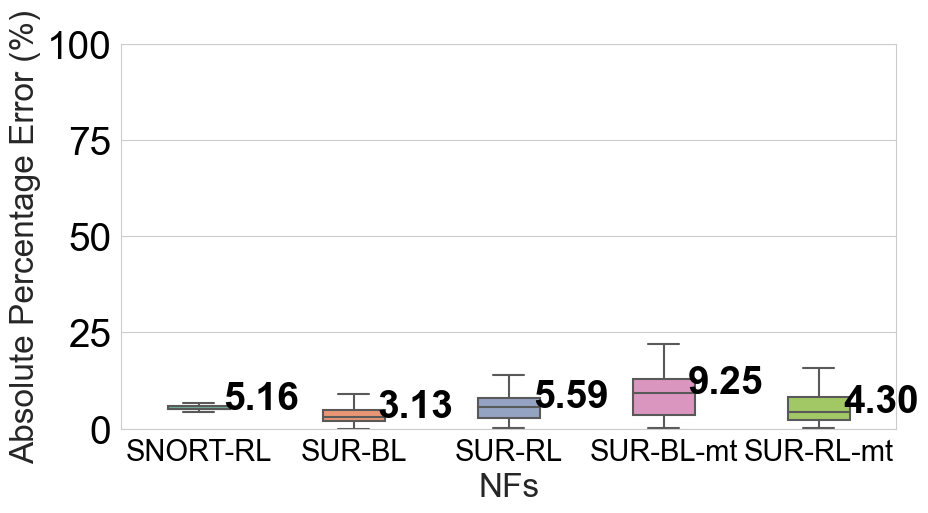}
       \caption{\label{fig:nfty-lite-internet-result}}
      \end{subfigure}
    \begin{subfigure}[b]{0.51\columnwidth}
    \centering
    \includegraphics[width=\columnwidth]{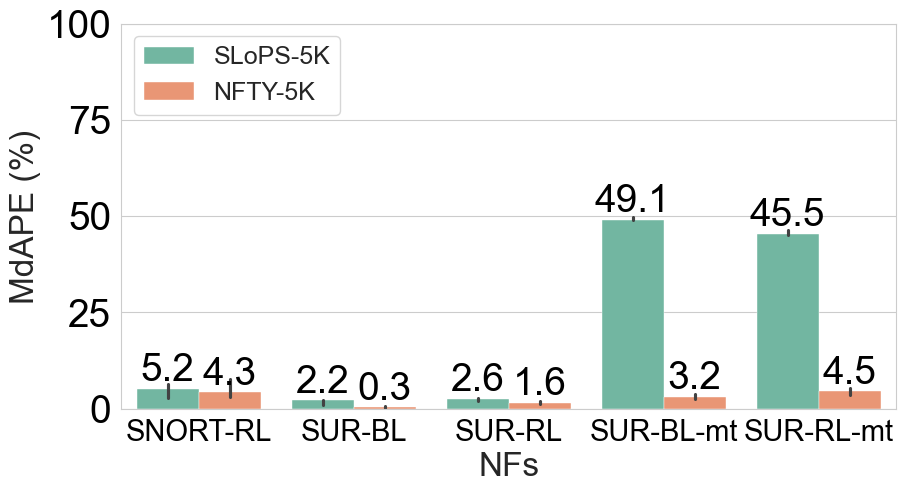}
    \caption{\label{fig:int-5k}}
\end{subfigure}
\begin{subfigure}[b]{0.51\columnwidth}
    \centering
    \includegraphics[width=\columnwidth]{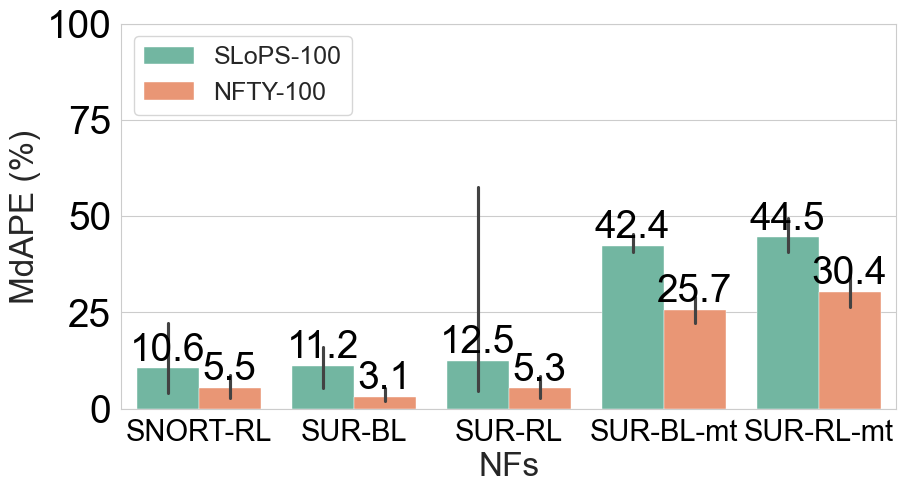}
    \caption{\label{fig:int-100}  }
\end{subfigure}
\caption{ (a) \syslarge results in 4\% to \syslargeinterneterror MdAPE in the Internet with optimized \NF deployment (with DVFS).(c) Comparing SLoPS-5K and \syslarge in internet experiments with an optimized sender and unoptimized \NF. (d) Same setting as (a), but now comparing NFTY-100 and SLoPS-100.}
\Description[\sys comparison with baseline in the Internet setting]{(a) Comparing SLoPS-5K and \syslarge in internet experiments with an optimized sender and unoptimized \NF. (b) Same setting as (a), but now comparing NFTY-100 and SLoPS-100.}
\end{figure*}

\subsection{Internet experiments}

For these experiments, we evaluate the case where packets traverse Internet2 and the public Internet \emph{after} being processed by the \NF, which is the worst-case scenario for the attacker. Indeed, any congestion that packets experience \emph{before} reaching the \NF, would not affect the \NF-induced dispersion.

\myitem{Setup:} For Internet2 results, our sender and the \NF node run in CloudLab Wisconsin, and our receiver node runs in Clemson as shown in Figure~\ref{fig:internet-setup}. The packets from the \NF to the receiver go through the Internet.  The sender and the receiver use the same nodes as in the controlled environment. The receiver uses Intel Haswell E5-2683 v3 nodes with 2.00 GHz 14-core CPUs.
We also evaluate NFTY in the public Internet, outside of the Internet2 setting, by hosting our receiver in  Vultr Cloud node in Los Angeles. Similar to Figure~\ref{fig:internet-setup}, as the packets traverse from the \NF to the Vultr machine, they traverse the public Internet.
We run these experiments at different times of the day to capture the effect of varying Internet congestion.

\myitem{Both \syslite and \syslarge remain within \sysliteinterneterror error in Internet2.}
 We evaluate \syslite (without DVFS at the \NF) in the Internet to show that \sys can accurately estimate the capacity even in the public Internet using only a small number of packets.  We summarize our results in Figure~\ref{fig:nfty-lite-internet-result}. \syslite can accurately estimate \NF's capacity within \sysliteinterneterror error. The spread in error values for faster \NFs is higher than in controlled experiments.
We also evaluate \syslarge (with DVFS at the \NF) to show that even with added noise from the Internet, shown in Figure~\ref{fig:internet-fs-results} the \syslarge's estimation using step detection works well.
\syslarge has a higher spread in MdAPE in the Internet compared to the controlled environment. However, the MdAPE in the Internet for \syslarge also remains within \syslargeinterneterror error.

\myitem{\syslarge remains within 8\% error in the public Internet, while \syslite is more affected by noise in the public Internet.} Figure ~\ref{fig:int-5k} shows that the MdAPE for NFTY-5k in this setting never exceeded 5\% for all \NFs. For all 100 experiments for each NF, the calculated error never exceeded 8\%. Thus, NFTY-5K achieves very low error rates even in the Internet. Figure~\ref{fig:int-100} indicates that NFTY-100 achieves low MdAPE (6\%) for the single-threaded \NFs. 
Compared to the spread of the Internet2 experiments, NFTY-100 has a greater spread  (not shown in figure~\ref{fig:int-100}) due to the added noise of the Internet.  For the multi-threaded \NFs, there are two queues that packets are processed in, and so each queue processes only 50 packets each. With so few dispersion values for each queue and the noise of the Internet prevalent at the beginning of a probe, NFTY produces an inaccurate estimate.
\myitem{Both NFTY-100 and NFTY-5K consistently outperform SLoPS.} When comparing performance of NFTY to SLoPS, it has a smaller MdAPE for all NFs for both techniques. As seen in the results for the controlled environment and Internet2, SLoPS performs poorly for the multi-threaded \NFs. 
Although NFTY-100 has reduced accuracy for the multi-threaded \NFs, it still outperforms SLoPS. When comparing performance for the single-threaded \NFs, NFTY-100 cuts the MdAPE by approximately 50\%. For NFTY-5K, it's performance is similar to SLoPS' for the single-threaded \NFs but reduces MdAPE by at least 90\% for the multi-threaded \NFs. 


\subsection{Case Study: \sys against Commercial NF in the Cloud} \label{sec:aws}

\begin{figure}[t!]
    \centering
    \begin{subfigure}[b]{0.48\columnwidth}
    \centering
    \includegraphics[width=\columnwidth]{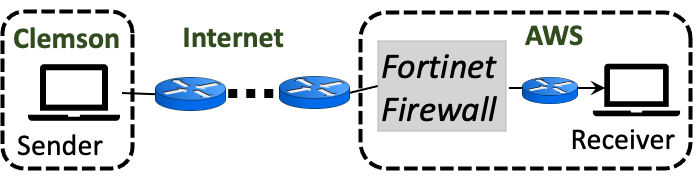}
    \caption{\label{fig:aws-setup}}
\end{subfigure}
\begin{subfigure}[b]{0.48\columnwidth}
    \centering
    \includegraphics[width=\columnwidth]{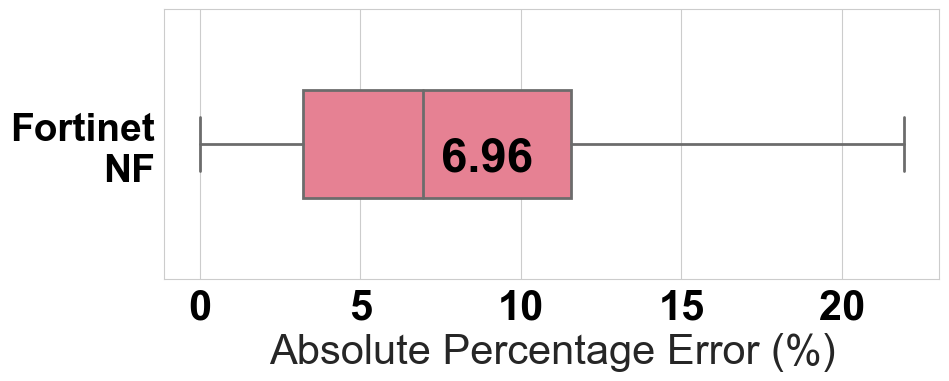}
    \caption{\label{fig:aws}  }
\end{subfigure}
\caption{ (a) In the AWS experiments,  the sender probes the \NF from the Internet. (b) \syslite results in \sysliteaws MdAPE while probing a commercial \NF Fortinet next generation firewall~\cite{fortinet}. The spread is high because here, the receiver and the NF are VMs and may induce extra noise.}
\Description[AWS Experiments]{(a) In the AWS experiments,  the sender probes the \NF from the Internet. (b) \syslite results in \sysliteaws MdAPE while probing a commercial \NF Fortinet next generation firewall~\cite{fortinet}. The spread is high because here, the receiver and the NF are VMs and may induce extra noise.}
\end{figure}
Having tested \sys in the lab and in the Internet we aim to investigate whether our results generalize beyond open-access \NFs and in-house deployments. To this end, we use \sys to calculate the capacity of the Fortinet next-generation firewall\cite{fortinet} which is a closed-source and commercial NF.

\myitem{Cloud setup:} We deploy the Fortinet firewall in the AWS US East cluster on an AWS c4.large instance type in front of a private subnet which was the recommended instance type by the vendor. Thus,  all traffic entering the subnet passes through the firewall as we show in Figure\ref{fig:aws-setup}. We have configured the firewall to use its default attack detection profiles for all UDP traffic. We find the ground truth processing capacity to be 62738 pps.
We measure the ground truth using a local node within AWS by flooding the NF and measuring the received rate at the receiver. We found that when we flood the NF, the receiver constantly receives packets at a rate of 62738 pps, and the remaining packets are dropped at the NF.
We use our Clemson Cloudlab node to probe the firewall using UDP traffic to implement \syslite. Observe that traffic from Clemson to the AWS nodes is forwarded through the public Internet and, thus is subject to cross-traffic.

We find that \syslite can accurately estimate the processing capacity of the Fortinet firewall within 7\% error. The spread in the error is high, which might be due to the the noise from use of a VM  of either \NF or the receiver node~\ref{fig:aws}. This shows that given an unknown NF with unknown deployment, \sys can accurately estimate its processing capacity.

\section{Countermeasures }



We present several countermeasures against an \NFCR attack, which essentially try to either conceal the dispersion values or the resources available at the \NF. We evaluate them for their effectiveness and the added overhead they impose. 

\myitem{Adding a random delay:} A network operator could add a random delay to packets, to affect the measured dispersion and thus obfuscate the real \NF processing delay. 
While intuitive and simple, its not clear whether this countermeasure will work in practice despite adding overhead to legitimate traffic since adding random delays might still preserve the average dispersion, making an \NFCR attack feasible. 

\myparatight{Additional packet batching:}
Rather than releasing packets as they arrive, an \NF could buffer and release them in batches to change the dispersion signature, and affect capacity estimation. 
The buffer size introduces a tradeoff between effectiveness and added latency to legitimate flows.

\myparatight{Reorder using the multiple-queues}
One could leverage multiple queues on NICs to trigger packet reordering. By forcing packets of the same flow to be sprayed across different queues, the attacker's packets would arrive reordered, effectively disrupting the dispersion signature. While this countermeasure would not delay traffic, it will affect the TCP performance which is known to suffer from reordering.

\myparatight{Rate-limiting}
Another alternative is to make the \NF appear to have lower capacity, by rate-limiting packets from the same flow or IP source for example. While an attacker mitigates this countermeasure by using multiple IPs, this would again require her to deal with packet re-orderings and also require keeping per-flow state which can be impractical. 
This countermeasure will have low overhead, as a single flow should not use \emph{even instantly} the full capacity of the \NF. 

\myparatight{Under-clocking the \NF}
An \NF can set the parameters of DVFS governors to stay at a lower frequency for longer and switch to higher frequencies after more sustained loads. 
As \sys sends a few thousand packets to trigger DVFS, with this countermeasure, an attacker would need to send many more packets making them detectable. 
However, delaying frequency scaling would process legitimate traffic slower. 

\begin{figure}[t]
    \centering
    \begin{subfigure}[b]{0.49\columnwidth}
    \centering
       \includegraphics[width=\columnwidth]{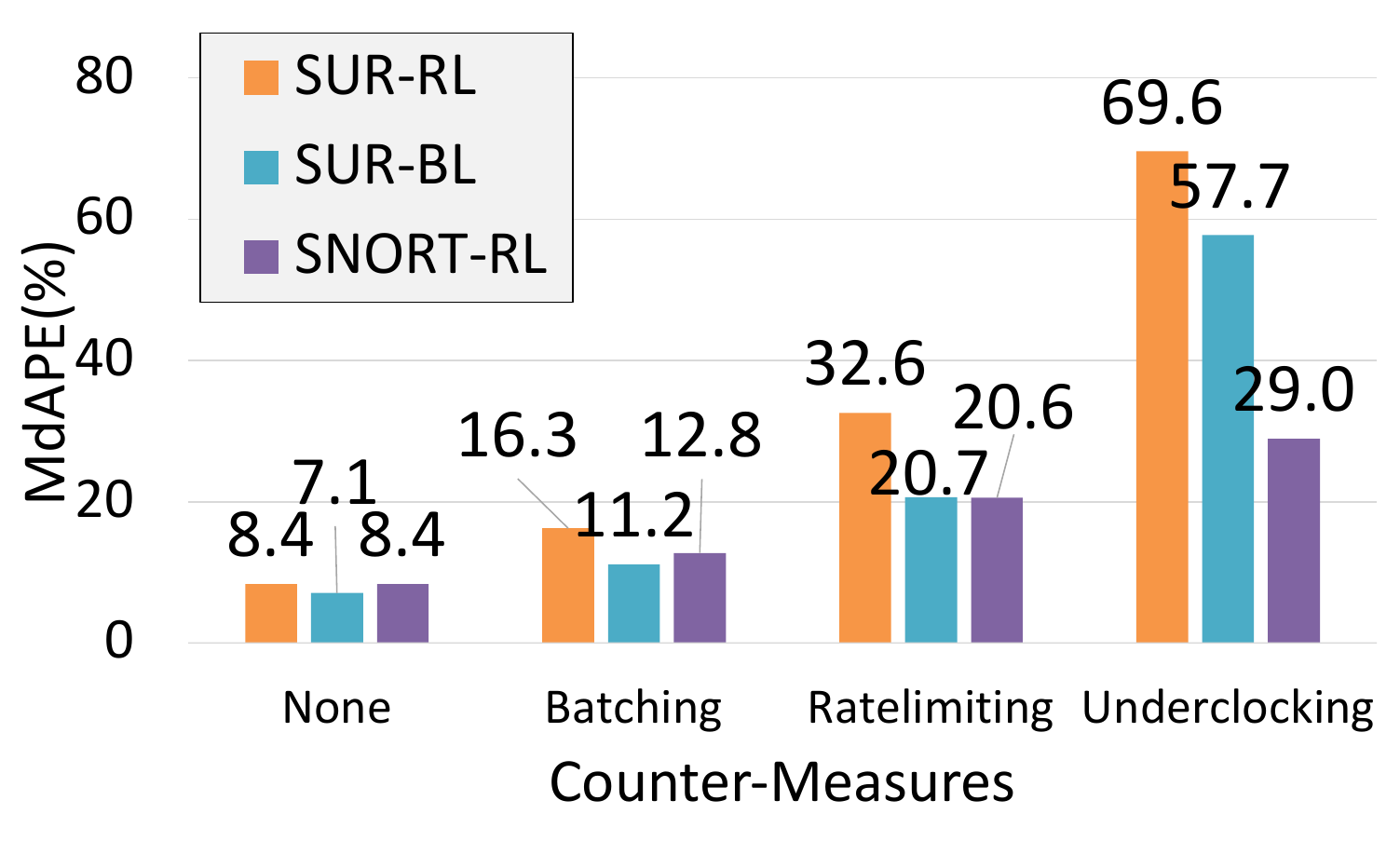}
    \caption{ \label{fig:counter-measures}}
\end{subfigure}
\begin{subfigure}[b]{0.49\columnwidth}
    \centering
    \includegraphics[width=\columnwidth]{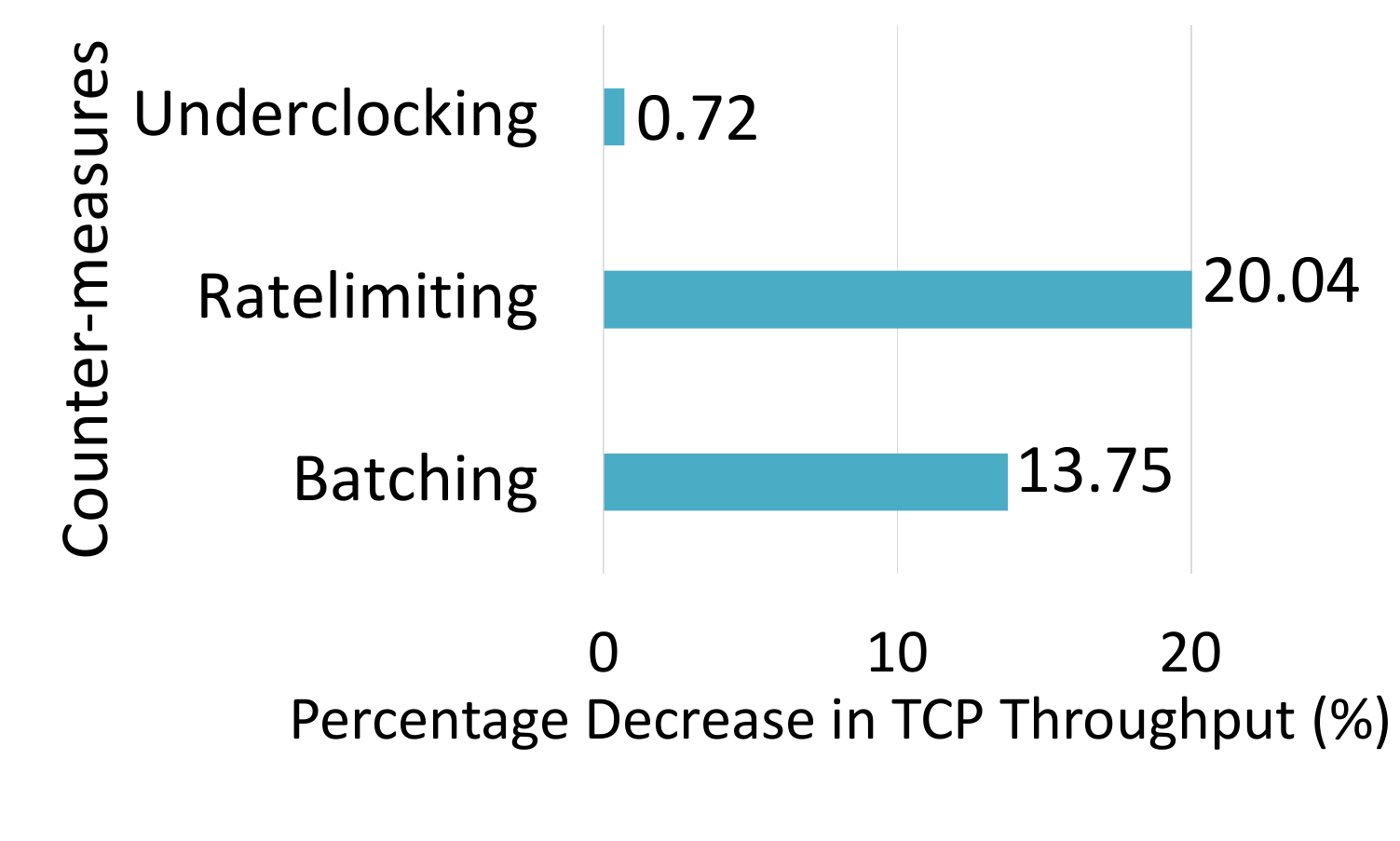}
    \caption{   \label{fig:cm-overhead} }
\end{subfigure}
\caption{ (a) \small Under-clocking is the most effective countermeasure. 
(b)\small The countermeasures impose overhead on legitimate traffic e.g.,  under-clocking would decrease the throughput of a TCP flow by 0.72\%.}
\Description[Performance of counter-measures]{Under-clocking is the most effective countermeasure. It would increase \sys's MAPE by 3x to 7x. (b)The countermeasures impose overhead on legitimate traffic. For example,  under-clocking would decrease the throughput if a TCP flow by 0.72\%.}
\vspace{-5pt}
\end{figure}

\myparatight{Countermeasures evaluation}
We investigate the trade-off between effectiveness and overhead of these countermeasures. 
To do so,  we measure the accuracy of \sys in estimating the capacity of three realistic \NFs, with various countermeasures deployed. To characterize overhead, we measure the loss in TCP throughput for a single Iperf~\cite{tirumala1999iperf} flow. We perform $10$ runs with each countermeasure and report the MdAPE and throughput reduction. 

We simulate three countermeasures, \textit{excessive batching} with a batch interval of $300\mu s$, \textit{rate-limiting} by 20\%, and \textit{under-clocking}. Batching aims at blurring the measured delay between packets (dispersion); rate-limiting permanently obfuscates the \NF's true capacity, while under-clocking hides it for some time.
We configure a batching interval of 500us, rate-limiting per flow to 20\% of the maximum bandwidth, and under-clocking by 100ms.

Figures~\ref{fig:counter-measures}, and \ref{fig:cm-overhead} summarize our results. We observe that under-clocking is the most effective countermeasure as it increases \sys MdAPE error by 3 to 5 times, while also incurring the least overhead since the low-frequency setting is only applicable for a short period.
Note that, this result is subject to our chosen overhead metric. Intuitively, the overhead of under-clocking would have been higher if we used another metric \eg tail flow completion time. 

Accurately evaluating the overhead of each trade-off is beyond the scope of our paper. Instead, this section provides various practical countermeasures that an operator could deploy based on their goals and the importance of their \NF. 

\section{Related Work}


In this section, we discuss related works beyond the link-bandwidth estimation, which we already compared with.


\myitem{Reconnaissance and prevention:} Prior work in network reconnaissance tries to infer the topology of the network to construct attacks~\cite{kang2013crossfire}. Similarly, Samak et~al.~\cite{samak2007firecracker} infer rules of a remote firewall by sending carefully crafted probes. Ramamurthy et~al.~\cite{ramamurthy2008remote} uses mini flash crowds to estimate various bottlenecks such as access bandwidth, CPU utilization, and memory usage of a server. Salehin \etal~\cite{salehin2013scheme} aims at estimating the delay in the networking stack of a server by sending packet probes. These works do not perform reconnaissance for \NF capacity and are thus orthogonal to our work. 
In addition to reconnaissance techniques, many prior works aim at protecting against reconnaissance attacks~\cite{jafarian2015effective,meier2018nethide,achleitner2017deceiving}. 

\myitem{\NF capacity modeling:} 
Multiple works aim at modeling \NFs for various reasons, 
\eg predicting memory impact on contention~\cite{manousis2020contention}, generating adversarial workloads~\cite{pedrosa2018automated}, or predicting performance~\cite{iyer2019performance}.
Such works are complementary to \NFCR and can be used to further optimize inference.

\section{Discussion}

\myitem{\sys measures the bottleneck \NF.} Note that in the case where there are multiple \NFs chained in a given network, \sys will only measure the packet processing capacity of the bottleneck \NF. To target a particular \NF, \sys would need one hop access to the \NF in the case that the \NF of interest is not the bottleneck \NF.

\myitem{\sys does not require the location of \NF.}
We propose \sys to estimate the packet processing capacity of an \NF deployed in a target network. Note that we do not know the location of the \NF. When \sys is run against a target network, it will automatically measure the packet processing capacity of the bottleneck \NF in the target network for a given packet type. Hence, \sys does not require explicit information about the location of the \NF. 

\myitem{\sys does not need the exact packet types of \NF.}
Since, the target of \sys is to measure the processing capacity of  bottlecneck \NFs for a given packet type, deployed in a target network, \sys can use popular coarse grained packet types \eg TCP SYN, UDP, DNS etc.,  to probe for packet processing capacity. Finding packet types that would lead to slow execution paths is an interesting future work direction. 

\myitem{\sys assumes links are not bottleneck.} Between the attacker and the \NF, \sys assumes that links are not the bottleneck, which would typically be true as \NFs do more computation than regular routers doing the forwarding. Note that for link bandwidth estimation techniques, they assume the alternate thing that \NFs are not the bottleneck. 

\myitem{\sys does not work for \NFs that elastically scale.} The current implementation of \sys does not work for \NFs that scale elastically with traffic load. Extending \sys to handle this scenario is possible, however, it will make \sys more noticeable, as it may need to trigger the elastic scaling. This in theory is very similar to DVFS, however, the number of packets  or the scale is very different. 
Hence, extending \sys to such \NFs while remaining stealthy is another interesting future work direction.

\myitem{\sys does not work for \NFs whose processing capacity change with packet history }
We scope \sys as a first step towards measuring the processing capacity of bottleneck \NFs remotely. In the future, \sys should be extended to include such stateful \NFs. 
To measure such an \NF, \sys would need to send multiple probes to figure out some part of the black box state machine of such an \NF. 

\myitem{\sys experiences trade-off between accuracy and detectability}
We evaluate \sys with two configurations. \syslite is less detectable but does not measure accurately in all cases. \syslarge is more visible but measures accurately in the cases we evaluated. We do not claim that \syslarge will be enough to detect accurately in all cases. In the end, it depends on the attacker's budget. Based on the attacker's budget, an attacker may measure accurately but get detected.
\section{Conclusion}

This paper presents the first formulation and feasibility analysis of Network Function Capacity Reconnaissance (\NFCR).
While anecdotal data suggests that attackers
have used some form of \NFCR to scout the resources
of their victims prior to actual DDoS attacks, such attempts are not documented, leaving network operators unable to detect or mitigate them.
To bridge this gap, we put ourselves in the shoes of an attacker and investigate various probing strategies, measurement infrastructures, and threat models. In doing so, we constructed a practical tool catered to the \NFCR problem, namely \sys. 
We identify and evaluate two representative \sys configurations, namely \syslite and \syslarge.  \syslite can accurately estimate the capacity of simple NF deployments while being extremely stealthy. \syslarge is accurate under more diverse \NF deployments but has a larger network footprint. 
While less accurate than \syslarge, \syslite is still powerful
enough to estimate the capacity of a commercial NF deployed
in AWS within 7\% error.
Finally, we present and evaluate countermeasures against \sys.


\appendix

\section{Ethics}
As with any ``attack'' paper, there is a risk that attackers can benefit from our work. This is largely outweighed by the benefit for benign operators that can better understand and prepare against such attacks. We have run our experiments on servers under our control, and the generated traffic is very small.



\bibliographystyle{ACM-Reference-Format}
\bibliography{cite}


\begin{thebibliography}{00}


\ifx \showCODEN    \undefined \def \showCODEN     #1{\unskip}     \fi
\ifx \showDOI      \undefined \def \showDOI       #1{#1}\fi
\ifx \showISBNx    \undefined \def \showISBNx     #1{\unskip}     \fi
\ifx \showISBNxiii \undefined \def \showISBNxiii  #1{\unskip}     \fi
\ifx \showISSN     \undefined \def \showISSN      #1{\unskip}     \fi
\ifx \showLCCN     \undefined \def \showLCCN      #1{\unskip}     \fi
\ifx \shownote     \undefined \def \shownote      #1{#1}          \fi
\ifx \showarticletitle \undefined \def \showarticletitle #1{#1}   \fi
\ifx \showURL      \undefined \def \showURL       {\relax}        \fi
\providecommand\bibfield[2]{#2}
\providecommand\bibinfo[2]{#2}
\providecommand\natexlab[1]{#1}
\providecommand\showeprint[2][]{arXiv:#2}

\bibitem[\protect\citeauthoryear{??}{jun}{2024}]%
        {juniper}
 \bibinfo{year}{2024}\natexlab{}.
\newblock   (\bibinfo{year}{2024}).
\newblock
\showURL{%
\url{https://www.juniper.net/documentation/us/en/software/junos/denial-of-service/topics/topic-map/security-network-dos-attack.html}}


\bibitem[\protect\citeauthoryear{Achleitner, La~Porta, McDaniel, Sugrim, Krishnamurthy, and Chadha}{Achleitner et~al\mbox{.}}{2017}]%
        {achleitner2017deceiving}
\bibfield{author}{\bibinfo{person}{Stefan Achleitner}, \bibinfo{person}{Thomas~F La~Porta}, \bibinfo{person}{Patrick McDaniel}, \bibinfo{person}{Shridatt Sugrim}, \bibinfo{person}{Srikanth~V Krishnamurthy}, {and} \bibinfo{person}{Ritu Chadha}.} \bibinfo{year}{2017}\natexlab{}.
\newblock \showarticletitle{Deceiving network reconnaissance using SDN-based virtual topologies}.
\newblock \bibinfo{journal}{{\em IEEE Transactions on Network and Service Management\/}} \bibinfo{volume}{14}, \bibinfo{number}{4} (\bibinfo{year}{2017}), \bibinfo{pages}{1098--1112}.
\newblock


\bibitem[\protect\citeauthoryear{Allman}{Allman}{2001}]%
        {allman2001measuring}
\bibfield{author}{\bibinfo{person}{Mark Allman}.} \bibinfo{year}{2001}\natexlab{}.
\newblock \showarticletitle{Measuring end-to-end bulk transfer capacity}.
\newblock  (\bibinfo{year}{2001}), \bibinfo{pages}{139--143}.
\newblock


\bibitem[\protect\citeauthoryear{Azure}{Azure}{2021}]%
        {azure-marketplace}
\bibfield{author}{\bibinfo{person}{Azure}.} \bibinfo{year}{2021}\natexlab{}.
\newblock \bibinfo{title}{Azure Marketplace}.
\newblock \bibinfo{howpublished}{\url{https://azuremarketplace.microsoft.com}}.   (\bibinfo{year}{2021}).
\newblock


\bibitem[\protect\citeauthoryear{Bai}{Bai}{1997}]%
        {bai1997estimating}
\bibfield{author}{\bibinfo{person}{Jushan Bai}.} \bibinfo{year}{1997}\natexlab{}.
\newblock \showarticletitle{Estimating multiple breaks one at a time}.
\newblock \bibinfo{journal}{{\em Econometric theory\/}} \bibinfo{volume}{13}, \bibinfo{number}{3} (\bibinfo{year}{1997}), \bibinfo{pages}{315--352}.
\newblock


\bibitem[\protect\citeauthoryear{Bolot}{Bolot}{1993}]%
        {bolot1993characterizing}
\bibfield{author}{\bibinfo{person}{Jean-Chrysostome Bolot}.} \bibinfo{year}{1993}\natexlab{}.
\newblock \showarticletitle{Characterizing end-to-end packet delay and loss in the internet}.
\newblock \bibinfo{journal}{{\em Journal of High Speed Networks\/}} \bibinfo{volume}{2}, \bibinfo{number}{3} (\bibinfo{year}{1993}), \bibinfo{pages}{305--323}.
\newblock


\bibitem[\protect\citeauthoryear{Carter and Crovella}{Carter and Crovella}{1996}]%
        {carter1996measuring}
\bibfield{author}{\bibinfo{person}{Robert~L Carter} {and} \bibinfo{person}{Mark~E Crovella}.} \bibinfo{year}{1996}\natexlab{}.
\newblock \showarticletitle{Measuring bottleneck link speed in packet-switched networks}.
\newblock \bibinfo{journal}{{\em Performance evaluation\/}}  \bibinfo{volume}{27} (\bibinfo{year}{1996}), \bibinfo{pages}{297--318}.
\newblock


\bibitem[\protect\citeauthoryear{Chen and Gupta}{Chen and Gupta}{2012}]%
        {chen2012parametric}
\bibfield{author}{\bibinfo{person}{Jie Chen} {and} \bibinfo{person}{Arjun~K Gupta}.} \bibinfo{year}{2012}\natexlab{}.
\newblock \bibinfo{booktitle}{{\em Parametric statistical change point analysis: with applications to genetics, medicine, and finance}}.
\newblock \bibinfo{publisher}{Springer}.
\newblock


\bibitem[\protect\citeauthoryear{Community}{Community}{2024}]%
        {snortcommunity_2024}
\bibfield{author}{\bibinfo{person}{Snort Community}.} \bibinfo{year}{2024}\natexlab{}.
\newblock \bibinfo{title}{What are Community Rules?}
\newblock   (\bibinfo{date}{May} \bibinfo{year}{2024}).
\newblock
\showURL{%
\url{https://www.snort.org/faq/what-are-community-rules}}


\bibitem[\protect\citeauthoryear{Corero}{Corero}{2014}]%
        {corero}
\bibfield{author}{\bibinfo{person}{Corero}.} \bibinfo{year}{2014}\natexlab{}.
\newblock \bibinfo{title}{Adaptive ddos attacks warrant next gen defense}.
\newblock \bibinfo{howpublished}{\url{https://informationsecuritybuzz.com/infosec-news/adaptive-ddos-attacks-warrant-next-gen-defense/}}.   (\bibinfo{year}{2014}).
\newblock


\bibitem[\protect\citeauthoryear{Detal, Hesmans, Bonaventure, Vanaubel, and Donnet}{Detal et~al\mbox{.}}{2013}]%
        {detal2013revealing}
\bibfield{author}{\bibinfo{person}{Gregory Detal}, \bibinfo{person}{Benjamin Hesmans}, \bibinfo{person}{Olivier Bonaventure}, \bibinfo{person}{Yves Vanaubel}, {and} \bibinfo{person}{Benoit Donnet}.} \bibinfo{year}{2013}\natexlab{}.
\newblock \showarticletitle{Revealing middlebox interference with tracebox}.
\newblock  (\bibinfo{year}{2013}), \bibinfo{pages}{1--8}.
\newblock


\bibitem[\protect\citeauthoryear{Dovrolis, Ramanathan, and Moore}{Dovrolis et~al\mbox{.}}{2001}]%
        {dovrolis2001packet}
\bibfield{author}{\bibinfo{person}{Constantinos Dovrolis}, \bibinfo{person}{Parameswaran Ramanathan}, {and} \bibinfo{person}{David Moore}.} \bibinfo{year}{2001}\natexlab{}.
\newblock \showarticletitle{What do packet dispersion techniques measure?}. In \bibinfo{booktitle}{{\em Proceedings IEEE INFOCOM 2001. Conference on Computer Communications. Twentieth Annual Joint Conference of the IEEE Computer and Communications Society (Cat. No. 01CH37213)}}, Vol.~\bibinfo{volume}{2}. \bibinfo{publisher}{IEEE}, \bibinfo{pages}{905--914}.
\newblock


\bibitem[\protect\citeauthoryear{Dovrolis, Ramanathan, and Moore}{Dovrolis et~al\mbox{.}}{2004}]%
        {dovrolis2004packet}
\bibfield{author}{\bibinfo{person}{Constantinos Dovrolis}, \bibinfo{person}{Parameswaran Ramanathan}, {and} \bibinfo{person}{David Moore}.} \bibinfo{year}{2004}\natexlab{}.
\newblock \showarticletitle{Packet-dispersion techniques and a capacity-estimation methodology}.
\newblock \bibinfo{journal}{{\em IEEE/ACM Transactions On Networking\/}} \bibinfo{volume}{12}, \bibinfo{number}{6} (\bibinfo{year}{2004}), \bibinfo{pages}{963--977}.
\newblock


\bibitem[\protect\citeauthoryear{Downey}{Downey}{1999}]%
        {clink}
\bibfield{author}{\bibinfo{person}{Allen~B. Downey}.} \bibinfo{year}{1999}\natexlab{}.
\newblock \showarticletitle{Using Pathchar to Estimate Internet Link Characteristics}.
\newblock \bibinfo{journal}{{\em SIGCOMM Comput. Commun. Rev.\/}} \bibinfo{volume}{29}, \bibinfo{number}{4} (\bibinfo{date}{aug} \bibinfo{year}{1999}), \bibinfo{pages}{241–250}.
\newblock
\showISSN{0146-4833}
\showDOI{%
\url{https://doi.org/10.1145/316194.316228}}


\bibitem[\protect\citeauthoryear{Duplyakin, Ricci, Maricq, Wong, Duerig, Eide, Stoller, Hibler, Johnson, Webb, et~al\mbox{.}}{Duplyakin et~al\mbox{.}}{2019}]%
        {duplyakin2019design}
\bibfield{author}{\bibinfo{person}{Dmitry Duplyakin}, \bibinfo{person}{Robert Ricci}, \bibinfo{person}{Aleksander Maricq}, \bibinfo{person}{Gary Wong}, \bibinfo{person}{Jonathon Duerig}, \bibinfo{person}{Eric Eide}, \bibinfo{person}{Leigh Stoller}, \bibinfo{person}{Mike Hibler}, \bibinfo{person}{David Johnson}, \bibinfo{person}{Kirk Webb}, {et~al\mbox{.}}} \bibinfo{year}{2019}\natexlab{}.
\newblock \showarticletitle{The Design and Operation of $\{$CloudLab$\}$}. In \bibinfo{booktitle}{{\em 2019 USENIX annual technical conference (USENIX ATC 19)}}. \bibinfo{publisher}{USENIX}, \bibinfo{pages}{1--14}.
\newblock


\bibitem[\protect\citeauthoryear{Fortinet}{Fortinet}{2023}]%
        {fortinet}
\bibfield{author}{\bibinfo{person}{Fortinet}.} \bibinfo{year}{2023}\natexlab{}.
\newblock \bibinfo{title}{Fortinet Cloud Security Amazon Web Services (AWS)}.
\newblock   (\bibinfo{year}{2023}).
\newblock
\showURL{%
\url{https://www.fortinet.com/products/public-cloud-security/aws}}


\bibitem[\protect\citeauthoryear{Fortinet}{Fortinet}{2024}]%
        {fortinet_2024}
\bibfield{author}{\bibinfo{person}{Fortinet}.} \bibinfo{year}{2024}\natexlab{}.
\newblock \bibinfo{title}{DoS Protection | FortiOS 6.4.15 Administration Guide}.
\newblock   (\bibinfo{date}{Apr} \bibinfo{year}{2024}).
\newblock
\showURL{%
\url{https://docs.fortinet.com/document/fortigate/6.4.15/administration-guide/771644/dos-protection}}


\bibitem[\protect\citeauthoryear{Fryzlewicz}{Fryzlewicz}{2014}]%
        {fryzlewicz2014wild}
\bibfield{author}{\bibinfo{person}{Piotr Fryzlewicz}.} \bibinfo{year}{2014}\natexlab{}.
\newblock \showarticletitle{Wild binary segmentation for multiple change-point detection}.
\newblock \bibinfo{journal}{{\em The Annals of Statistics\/}} \bibinfo{volume}{42}, \bibinfo{number}{6} (\bibinfo{year}{2014}), \bibinfo{pages}{2243--2281}.
\newblock


\bibitem[\protect\citeauthoryear{Garreau and Arlot}{Garreau and Arlot}{2018}]%
        {garreau2018consistent}
\bibfield{author}{\bibinfo{person}{Damien Garreau} {and} \bibinfo{person}{Sylvain Arlot}.} \bibinfo{year}{2018}\natexlab{}.
\newblock \showarticletitle{Consistent change-point detection with kernels}.
\newblock \bibinfo{journal}{{\em Electronic Journal of Statistics\/}} \bibinfo{volume}{12}, \bibinfo{number}{2} (\bibinfo{year}{2018}), \bibinfo{pages}{4440--4486}.
\newblock


\bibitem[\protect\citeauthoryear{Gunes and Sarac}{Gunes and Sarac}{2008}]%
        {gunes2008resolving}
\bibfield{author}{\bibinfo{person}{Mehmet~Hadi Gunes} {and} \bibinfo{person}{Kamil Sarac}.} \bibinfo{year}{2008}\natexlab{}.
\newblock \showarticletitle{Resolving anonymous routers in internet topology measurement studies}. In \bibinfo{booktitle}{{\em IEEE INFOCOM 2008-The 27th Conference on Computer Communications}}. IEEE, \bibinfo{publisher}{IEEE}, \bibinfo{pages}{1076--1084}.
\newblock


\bibitem[\protect\citeauthoryear{IBM}{IBM}{2021}]%
        {IC}
\bibfield{author}{\bibinfo{person}{IBM}.} \bibinfo{year}{2021}\natexlab{}.
\newblock \bibinfo{title}{{{Interrupt Coalescing }}}.
\newblock \bibinfo{howpublished}{\url{https://prod.ibmdocs-production-dal-6099123ce774e592a519d7c33db8265e-0000.us-south.containers.appdomain.cloud/docs/en/aix/7.2?topic=options-interrupt-coalescing}}.   (\bibinfo{date}{Oct.} \bibinfo{year}{2021}).
\newblock


\bibitem[\protect\citeauthoryear{Intel}{Intel}{2023}]%
        {ReceiveSideScaling}
\bibfield{author}{\bibinfo{person}{Intel}.} \bibinfo{year}{2023}\natexlab{}.
\newblock \bibinfo{title}{Receive {{Side Scaling}} on {{Intel}}\textregistered{} {{Network Adapters}}}.
\newblock \bibinfo{howpublished}{\url{https://www.intel.com/content/www/us/en/support/articles/000006703/network-and-i-o/ethernet-products.html}}.   (\bibinfo{year}{2023}).
\newblock


\bibitem[\protect\citeauthoryear{Iyer, Pedrosa, Zaostrovnykh, Pirelli, Argyraki, and Candea}{Iyer et~al\mbox{.}}{2019}]%
        {iyer2019performance}
\bibfield{author}{\bibinfo{person}{Rishabh Iyer}, \bibinfo{person}{Luis Pedrosa}, \bibinfo{person}{Arseniy Zaostrovnykh}, \bibinfo{person}{Solal Pirelli}, \bibinfo{person}{Katerina Argyraki}, {and} \bibinfo{person}{George Candea}.} \bibinfo{year}{2019}\natexlab{}.
\newblock \showarticletitle{Performance contracts for software network functions}. In \bibinfo{booktitle}{{\em 16th USENIX Symposium on Networked Systems Design and Implementation (NSDI 19)}}. \bibinfo{publisher}{USENIX}, \bibinfo{pages}{517--530}.
\newblock


\bibitem[\protect\citeauthoryear{Jacobson}{Jacobson}{1997}]%
        {jacobson1997pathchar}
\bibfield{author}{\bibinfo{person}{Van Jacobson}.} \bibinfo{year}{1997}\natexlab{}.
\newblock \bibinfo{title}{Pathchar: A tool to infer characteristics of Internet paths}.
\newblock   (\bibinfo{year}{1997}).
\newblock


\bibitem[\protect\citeauthoryear{Jafarian, Al-Shaer, and Duan}{Jafarian et~al\mbox{.}}{2015}]%
        {jafarian2015effective}
\bibfield{author}{\bibinfo{person}{Jafar~Haadi Jafarian}, \bibinfo{person}{Ehab Al-Shaer}, {and} \bibinfo{person}{Qi Duan}.} \bibinfo{year}{2015}\natexlab{}.
\newblock \showarticletitle{An effective address mutation approach for disrupting reconnaissance attacks}.
\newblock \bibinfo{journal}{{\em IEEE Transactions on Information Forensics and Security\/}} \bibinfo{volume}{10}, \bibinfo{number}{12} (\bibinfo{year}{2015}), \bibinfo{pages}{2562--2577}.
\newblock


\bibitem[\protect\citeauthoryear{Jain and Dovrolis}{Jain and Dovrolis}{2002}]%
        {jain2002end}
\bibfield{author}{\bibinfo{person}{Manish Jain} {and} \bibinfo{person}{Constantinos Dovrolis}.} \bibinfo{year}{2002}\natexlab{}.
\newblock \showarticletitle{End-to-end available bandwidth: Measurement methodology, dynamics, and relation with TCP throughput}.
\newblock \bibinfo{journal}{{\em ACM SIGCOMM Computer Communication Review\/}} \bibinfo{volume}{32}, \bibinfo{number}{4} (\bibinfo{year}{2002}), \bibinfo{pages}{295--308}.
\newblock


\bibitem[\protect\citeauthoryear{Johnsson, Melander, Bj{\"o}rkman, and Bjorkman}{Johnsson et~al\mbox{.}}{2004}]%
        {johnsson2004diettopp}
\bibfield{author}{\bibinfo{person}{Andreas Johnsson}, \bibinfo{person}{Bob Melander}, \bibinfo{person}{Mats Bj{\"o}rkman}, {and} \bibinfo{person}{M Bjorkman}.} \bibinfo{year}{2004}\natexlab{}.
\newblock \showarticletitle{Diettopp: A first implementation and evaluation of a simplified bandwidth measurement method}. In \bibinfo{booktitle}{{\em Second Swedish National Computer Networking Workshop}}, Vol.~\bibinfo{volume}{5}. Citeseer.
\newblock


\bibitem[\protect\citeauthoryear{Kachan, Siemens, and Shuvalov}{Kachan et~al\mbox{.}}{2015}]%
        {kachan2015available}
\bibfield{author}{\bibinfo{person}{Dmitry Kachan}, \bibinfo{person}{Eduard Siemens}, {and} \bibinfo{person}{Vyacheslav Shuvalov}.} \bibinfo{year}{2015}\natexlab{}.
\newblock \showarticletitle{Available bandwidth measurement for 10 Gbps networks}. In \bibinfo{booktitle}{{\em 2015 International Siberian Conference on Control and Communications (SIBCON)}}. IEEE, \bibinfo{pages}{1--10}.
\newblock


\bibitem[\protect\citeauthoryear{Kang, Lee, and Gligor}{Kang et~al\mbox{.}}{2013}]%
        {kang2013crossfire}
\bibfield{author}{\bibinfo{person}{Min~Suk Kang}, \bibinfo{person}{Soo~Bum Lee}, {and} \bibinfo{person}{Virgil~D Gligor}.} \bibinfo{year}{2013}\natexlab{}.
\newblock \showarticletitle{The crossfire attack}. In \bibinfo{booktitle}{{\em 2013 IEEE symposium on security and privacy}}. IEEE, \bibinfo{publisher}{IEEE}, \bibinfo{pages}{127--141}.
\newblock


\bibitem[\protect\citeauthoryear{Kapoor, Chen, Lao, Gerla, and Sanadidi}{Kapoor et~al\mbox{.}}{2004}]%
        {kapoor2004capprobe}
\bibfield{author}{\bibinfo{person}{Rohit Kapoor}, \bibinfo{person}{Ling-Jyh Chen}, \bibinfo{person}{Li Lao}, \bibinfo{person}{Mario Gerla}, {and} \bibinfo{person}{M~Young Sanadidi}.} \bibinfo{year}{2004}\natexlab{}.
\newblock \showarticletitle{Capprobe: A simple and accurate capacity estimation technique}.
\newblock \bibinfo{journal}{{\em ACM SIGCOMM Computer Communication Review\/}} \bibinfo{volume}{34}, \bibinfo{number}{4} (\bibinfo{year}{2004}), \bibinfo{pages}{67--78}.
\newblock


\bibitem[\protect\citeauthoryear{Kohler, Morris, Chen, Jannotti, and Kaashoek}{Kohler et~al\mbox{.}}{2000}]%
        {kohler2000click}
\bibfield{author}{\bibinfo{person}{Eddie Kohler}, \bibinfo{person}{Robert Morris}, \bibinfo{person}{Benjie Chen}, \bibinfo{person}{John Jannotti}, {and} \bibinfo{person}{M~Frans Kaashoek}.} \bibinfo{year}{2000}\natexlab{}.
\newblock \showarticletitle{The Click modular router}.
\newblock \bibinfo{journal}{{\em ACM Transactions on Computer Systems (TOCS)\/}} \bibinfo{volume}{18}, \bibinfo{number}{3} (\bibinfo{year}{2000}), \bibinfo{pages}{263--297}.
\newblock


\bibitem[\protect\citeauthoryear{Lai and Baker}{Lai and Baker}{1999}]%
        {lai1999measuring}
\bibfield{author}{\bibinfo{person}{Kevin Lai} {and} \bibinfo{person}{Mary Baker}.} \bibinfo{year}{1999}\natexlab{}.
\newblock \showarticletitle{Measuring bandwidth}. In \bibinfo{booktitle}{{\em IEEE INFOCOM'99. Conference on Computer Communications. Proceedings. Eighteenth Annual Joint Conference of the IEEE Computer and Communications Societies. The Future is Now (Cat. No. 99CH36320)}}, Vol.~\bibinfo{volume}{1}. IEEE, \bibinfo{publisher}{IEEE}, \bibinfo{pages}{235--245}.
\newblock


\bibitem[\protect\citeauthoryear{Lai and Baker}{Lai and Baker}{2001}]%
        {lai2001nettimer}
\bibfield{author}{\bibinfo{person}{Kevin Lai} {and} \bibinfo{person}{Mary Baker}.} \bibinfo{year}{2001}\natexlab{}.
\newblock \showarticletitle{Nettimer: A tool for measuring bottleneck link bandwidth}. In \bibinfo{booktitle}{{\em 3rd USENIX Symposium on Internet Technologies and Systems (USITS 01)}}.
\newblock


\bibitem[\protect\citeauthoryear{Li, Claypool, and Kinicki}{Li et~al\mbox{.}}{2008}]%
        {li2008wbest}
\bibfield{author}{\bibinfo{person}{Mingzhe Li}, \bibinfo{person}{Mark Claypool}, {and} \bibinfo{person}{Robert Kinicki}.} \bibinfo{year}{2008}\natexlab{}.
\newblock \showarticletitle{WBest: A bandwidth estimation tool for IEEE 802.11 wireless networks}. In \bibinfo{booktitle}{{\em 2008 33rd IEEE Conference on Local Computer Networks (LCN)}}. IEEE, \bibinfo{publisher}{IEEE}, \bibinfo{pages}{374--381}.
\newblock


\bibitem[\protect\citeauthoryear{Liang}{Liang}{2017}]%
        {liang_2017}
\bibfield{author}{\bibinfo{person}{Kan Liang}.} \bibinfo{year}{2017}\natexlab{}.
\newblock \bibinfo{title}{Improve network performance by setting per-queue interrupt moderation in Linux*}.
\newblock   (\bibinfo{date}{May} \bibinfo{year}{2017}).
\newblock
\showURL{%
\url{https://01.org/linux-interrupt-moderation}}


\bibitem[\protect\citeauthoryear{Man}{Man}{2023}]%
        {TracerouteLinuxMan}
\bibfield{author}{\bibinfo{person}{Linux Man}.} \bibinfo{year}{2023}\natexlab{}.
\newblock \bibinfo{title}{Traceroute(8) - {{Linux}} Man Page}.
\newblock \bibinfo{howpublished}{\url{https://linux.die.net/man/8/traceroute}}.   (\bibinfo{year}{2023}).
\newblock


\bibitem[\protect\citeauthoryear{Manousis, Sharma, Sekar, and Sherry}{Manousis et~al\mbox{.}}{2020}]%
        {manousis2020contention}
\bibfield{author}{\bibinfo{person}{Antonis Manousis}, \bibinfo{person}{Rahul~Anand Sharma}, \bibinfo{person}{Vyas Sekar}, {and} \bibinfo{person}{Justine Sherry}.} \bibinfo{year}{2020}\natexlab{}.
\newblock \showarticletitle{Contention-aware performance prediction for virtualized network functions}. In \bibinfo{booktitle}{{\em Proceedings of the Annual conference of the ACM Special Interest Group on Data Communication on the applications, technologies, architectures, and protocols for computer communication}}. \bibinfo{publisher}{ACM}, \bibinfo{pages}{270--282}.
\newblock


\bibitem[\protect\citeauthoryear{Mathis}{Mathis}{1996}]%
        {mathis1996diagnosing}
\bibfield{author}{\bibinfo{person}{Matthew Mathis}.} \bibinfo{year}{1996}\natexlab{}.
\newblock \showarticletitle{Diagnosing internet congestion with a transport layer performance tool}.
\newblock \bibinfo{journal}{{\em Proc. INET'96, June\/}} (\bibinfo{year}{1996}).
\newblock


\bibitem[\protect\citeauthoryear{Meier, Tsankov, Lenders, Vanbever, and Vechev}{Meier et~al\mbox{.}}{2018}]%
        {meier2018nethide}
\bibfield{author}{\bibinfo{person}{Roland Meier}, \bibinfo{person}{Petar Tsankov}, \bibinfo{person}{Vincent Lenders}, \bibinfo{person}{Laurent Vanbever}, {and} \bibinfo{person}{Martin Vechev}.} \bibinfo{year}{2018}\natexlab{}.
\newblock \showarticletitle{$\{$NetHide$\}$: Secure and Practical Network Topology Obfuscation}. In \bibinfo{booktitle}{{\em 27th USENIX Security Symposium (USENIX Security 18)}}. \bibinfo{publisher}{USENIX}, \bibinfo{pages}{693--709}.
\newblock


\bibitem[\protect\citeauthoryear{Networks}{Networks}{2024}]%
        {panw_2024}
\bibfield{author}{\bibinfo{person}{Palo~Alto Networks}.} \bibinfo{year}{2024}\natexlab{}.
\newblock \bibinfo{title}{Flood Protection}.
\newblock   (\bibinfo{date}{Jan} \bibinfo{year}{2024}).
\newblock
\showURL{%
\url{https://docs.paloaltonetworks.com/pan-os/9-1/pan-os-web-interface-help/network/network-network-profiles/network-network-profiles-zone-protection/flood-protection}}


\bibitem[\protect\citeauthoryear{Panjwani, Tan, Jarrin, and Cukier}{Panjwani et~al\mbox{.}}{2005}]%
        {panjwani2005experimental}
\bibfield{author}{\bibinfo{person}{Susmit Panjwani}, \bibinfo{person}{Stephanie Tan}, \bibinfo{person}{Keith~M Jarrin}, {and} \bibinfo{person}{Michel Cukier}.} \bibinfo{year}{2005}\natexlab{}.
\newblock \showarticletitle{An experimental evaluation to determine if port scans are precursors to an attack}. In \bibinfo{booktitle}{{\em 2005 International Conference on Dependable Systems and Networks (DSN'05)}}. IEEE, \bibinfo{publisher}{IEEE}, \bibinfo{pages}{602--611}.
\newblock


\bibitem[\protect\citeauthoryear{Paxson}{Paxson}{1997}]%
        {paxson1997end}
\bibfield{author}{\bibinfo{person}{Vern Paxson}.} \bibinfo{year}{1997}\natexlab{}.
\newblock \showarticletitle{End-to-end Internet packet dynamics}. In \bibinfo{booktitle}{{\em Proceedings of the ACM SIGCOMM'97 conference on Applications, technologies, architectures, and protocols for computer communication}}. \bibinfo{publisher}{ACM}, \bibinfo{pages}{139--152}.
\newblock


\bibitem[\protect\citeauthoryear{Pedrosa, Iyer, Zaostrovnykh, Fietz, and Argyraki}{Pedrosa et~al\mbox{.}}{2018}]%
        {pedrosa2018automated}
\bibfield{author}{\bibinfo{person}{Luis Pedrosa}, \bibinfo{person}{Rishabh Iyer}, \bibinfo{person}{Arseniy Zaostrovnykh}, \bibinfo{person}{Jonas Fietz}, {and} \bibinfo{person}{Katerina Argyraki}.} \bibinfo{year}{2018}\natexlab{}.
\newblock \showarticletitle{Automated synthesis of adversarial workloads for network functions}. In \bibinfo{booktitle}{{\em Proceedings of the 2018 Conference of the ACM Special Interest Group on Data Communication}}. \bibinfo{publisher}{ACM}, \bibinfo{pages}{372--385}.
\newblock


\bibitem[\protect\citeauthoryear{Prasad, Jain, and Dovrolis}{Prasad et~al\mbox{.}}{2004}]%
        {prasad2004effects}
\bibfield{author}{\bibinfo{person}{Ravi Prasad}, \bibinfo{person}{Manish Jain}, {and} \bibinfo{person}{Constantinos Dovrolis}.} \bibinfo{year}{2004}\natexlab{}.
\newblock \showarticletitle{Effects of interrupt coalescence on network measurements}. In \bibinfo{booktitle}{{\em International Workshop on Passive and Active Network Measurement}}. Springer, \bibinfo{publisher}{Springer}, \bibinfo{pages}{247--256}.
\newblock


\bibitem[\protect\citeauthoryear{Ramamurthy, Sekar, Akella, Krishnamurthy, and Shaikh}{Ramamurthy et~al\mbox{.}}{2008}]%
        {ramamurthy2008remote}
\bibfield{author}{\bibinfo{person}{Pratap Ramamurthy}, \bibinfo{person}{Vyas Sekar}, \bibinfo{person}{Aditya Akella}, \bibinfo{person}{Balachander Krishnamurthy}, {and} \bibinfo{person}{Anees Shaikh}.} \bibinfo{year}{2008}\natexlab{}.
\newblock \showarticletitle{Remote Profiling of Resource Constraints of Web Servers Using Mini-Flash Crowds.} \bibinfo{publisher}{USENIX}.
\newblock


\bibitem[\protect\citeauthoryear{Ravaioli, Urvoy-Keller, and Barakat}{Ravaioli et~al\mbox{.}}{2015}]%
        {ravaioli2015characterizing}
\bibfield{author}{\bibinfo{person}{Riccardo Ravaioli}, \bibinfo{person}{Guillaume Urvoy-Keller}, {and} \bibinfo{person}{Chadi Barakat}.} \bibinfo{year}{2015}\natexlab{}.
\newblock \showarticletitle{Characterizing ICMP rate limitation on routers}. In \bibinfo{booktitle}{{\em 2015 IEEE International Conference on Communications (ICC)}}. IEEE, \bibinfo{publisher}{IEEE}, \bibinfo{pages}{6043--6049}.
\newblock


\bibitem[\protect\citeauthoryear{Ribeiro, Riedi, Baraniuk, Navratil, and Cottrell}{Ribeiro et~al\mbox{.}}{2003}]%
        {ribeiro2003pathchirp}
\bibfield{author}{\bibinfo{person}{Vinay~Joseph Ribeiro}, \bibinfo{person}{Rudolf~H Riedi}, \bibinfo{person}{Richard~G Baraniuk}, \bibinfo{person}{Jiri Navratil}, {and} \bibinfo{person}{Les Cottrell}.} \bibinfo{year}{2003}\natexlab{}.
\newblock \showarticletitle{pathchirp: Efficient available bandwidth estimation for network paths}. In \bibinfo{booktitle}{{\em Passive and active measurement workshop}}. \bibinfo{publisher}{Springer}.
\newblock


\bibitem[\protect\citeauthoryear{Rohith, Moharir, Shobha, et~al\mbox{.}}{Rohith et~al\mbox{.}}{2018}]%
        {rohith2018scapy}
\bibfield{author}{\bibinfo{person}{R Rohith}, \bibinfo{person}{Minal Moharir}, \bibinfo{person}{G Shobha}, {et~al\mbox{.}}} \bibinfo{year}{2018}\natexlab{}.
\newblock \showarticletitle{SCAPY-A powerful interactive packet manipulation program}. In \bibinfo{booktitle}{{\em 2018 international conference on networking, embedded and wireless systems (ICNEWS)}}. IEEE, \bibinfo{publisher}{IEEE}, \bibinfo{pages}{1--5}.
\newblock


\bibitem[\protect\citeauthoryear{Salehin, Rojas-Cessa, Lin, Dong, and Kijkanjanarat}{Salehin et~al\mbox{.}}{2013}]%
        {salehin2013scheme}
\bibfield{author}{\bibinfo{person}{Khondaker~M Salehin}, \bibinfo{person}{Roberto Rojas-Cessa}, \bibinfo{person}{Chuan-bi Lin}, \bibinfo{person}{Ziqian Dong}, {and} \bibinfo{person}{Taweesak Kijkanjanarat}.} \bibinfo{year}{2013}\natexlab{}.
\newblock \showarticletitle{Scheme to measure packet processing time of a remote host through estimation of end-link capacity}.
\newblock \bibinfo{journal}{{\it IEEE Trans. Comput.}} \bibinfo{volume}{64}, \bibinfo{number}{1} (\bibinfo{year}{2013}), \bibinfo{pages}{205--218}.
\newblock


\bibitem[\protect\citeauthoryear{Salim, Olsson, and Kuznetsov}{Salim et~al\mbox{.}}{2001}]%
        {salim2001beyond}
\bibfield{author}{\bibinfo{person}{Jamal~Hadi Salim}, \bibinfo{person}{Robert Olsson}, {and} \bibinfo{person}{Alexey Kuznetsov}.} \bibinfo{year}{2001}\natexlab{}.
\newblock \showarticletitle{Beyond softnet}. In \bibinfo{booktitle}{{\em 5th Annual Linux Showcase \& Conference (ALS 01)}}.
\newblock


\bibitem[\protect\citeauthoryear{Samak, El-Atawy, and Al-Shaer}{Samak et~al\mbox{.}}{2007}]%
        {samak2007firecracker}
\bibfield{author}{\bibinfo{person}{Taghrid Samak}, \bibinfo{person}{Adel El-Atawy}, {and} \bibinfo{person}{Ehab Al-Shaer}.} \bibinfo{year}{2007}\natexlab{}.
\newblock \showarticletitle{Firecracker: A framework for inferring firewall policies using smart probing}. In \bibinfo{booktitle}{{\em 2007 IEEE International Conference on Network Protocols}}. IEEE, \bibinfo{publisher}{IEEE}, \bibinfo{pages}{294--303}.
\newblock


\bibitem[\protect\citeauthoryear{Saroiu, Gummadi, and Gribble}{Saroiu et~al\mbox{.}}{2002}]%
        {saroiu2002sprobe}
\bibfield{author}{\bibinfo{person}{Stefan Saroiu}, \bibinfo{person}{P~Krishna Gummadi}, {and} \bibinfo{person}{Steven~D Gribble}.} \bibinfo{year}{2002}\natexlab{}.
\newblock \showarticletitle{Sprobe: A fast technique for measuring bottleneck bandwidth in uncooperative environments}. In \bibinfo{booktitle}{{\em IEEE INFOCOM}}. \bibinfo{publisher}{IEEE}, \bibinfo{pages}{1}.
\newblock


\bibitem[\protect\citeauthoryear{Schneier}{Schneier}{2016}]%
        {schneier_2016}
\bibfield{author}{\bibinfo{person}{Bruce Schneier}.} \bibinfo{year}{2016}\natexlab{}.
\newblock \bibinfo{title}{Someone Is Learning How to Take Down the Internet}.
\newblock   (\bibinfo{date}{Sep} \bibinfo{year}{2016}).
\newblock
\showURL{%
\url{https://www.schneier.com/blog/archives/2016/09/someone_is_lear.html}}


\bibitem[\protect\citeauthoryear{Snort}{Snort}{2023}]%
        {SnortNetworkIntrusion}
\bibfield{author}{\bibinfo{person}{Snort}.} \bibinfo{year}{2023}\natexlab{}.
\newblock \bibinfo{title}{Snort - {{Network Intrusion Detection}} \& {{Prevention System}}}.
\newblock \bibinfo{howpublished}{\url{https://www.snort.org/}}.   (\bibinfo{year}{2023}).
\newblock


\bibitem[\protect\citeauthoryear{Suricata}{Suricata}{2023}]%
        {Suricata}
\bibfield{author}{\bibinfo{person}{Suricata}.} \bibinfo{year}{2023}\natexlab{}.
\newblock \bibinfo{title}{Suricata}.
\newblock \bibinfo{howpublished}{\url{https://suricata.io/}}.   (\bibinfo{year}{2023}).
\newblock


\bibitem[\protect\citeauthoryear{TechHub}{TechHub}{2023}]%
        {ICMPRatelimiting}
\bibfield{author}{\bibinfo{person}{TechHub}.} \bibinfo{year}{2023}\natexlab{}.
\newblock \bibinfo{title}{{{ICMP}} Rate-Limiting}.
\newblock \bibinfo{howpublished}{\url{https://techhub.hpe.com/eginfolib/networking/docs/switches/RA/15-18/5998-8161\_ra\_2620\_mcg/content/ch05s03.html}}.   (\bibinfo{year}{2023}).
\newblock


\bibitem[\protect\citeauthoryear{Threats}{Threats}{2023}]%
        {ProofpointEmergingThreats}
\bibfield{author}{\bibinfo{person}{Proofpoint~Emerging Threats}.} \bibinfo{year}{2023}\natexlab{}.
\newblock \bibinfo{title}{Proofpoint {{Emerging Threats Rules}}}.
\newblock \bibinfo{howpublished}{\url{https://rules.emergingthreats.net/open/suricata/rules/}}.   (\bibinfo{year}{2023}).
\newblock


\bibitem[\protect\citeauthoryear{Tirumala}{Tirumala}{1999}]%
        {tirumala1999iperf}
\bibfield{author}{\bibinfo{person}{Ajay Tirumala}.} \bibinfo{year}{1999}\natexlab{}.
\newblock \showarticletitle{Iperf: The TCP/UDP bandwidth measurement tool}.
\newblock \bibinfo{journal}{{\em http://dast. nlanr. net/Projects/Iperf/\/}} (\bibinfo{year}{1999}).
\newblock


\bibitem[\protect\citeauthoryear{Truong, Oudre, and Vayatis}{Truong et~al\mbox{.}}{2020}]%
        {truong2020selective}
\bibfield{author}{\bibinfo{person}{Charles Truong}, \bibinfo{person}{Laurent Oudre}, {and} \bibinfo{person}{Nicolas Vayatis}.} \bibinfo{year}{2020}\natexlab{}.
\newblock \showarticletitle{Selective review of offline change point detection methods}.
\newblock \bibinfo{journal}{{\em Signal Processing\/}}  \bibinfo{volume}{167} (\bibinfo{year}{2020}), \bibinfo{pages}{107299}.
\newblock


\bibitem[\protect\citeauthoryear{verisign}{verisign}{2016}]%
        {verisign_blog_2016}
verisign \bibinfo{year}{2016}\natexlab{}.
\newblock \bibinfo{title}{VERISIGN Q2 2016 DDOS TRENDS: LAYER 7 DDOS ATTACKS A GROWING TREND}.
\newblock \bibinfo{howpublished}{\url{https://blog.verisign.com/security/verisign-q2-2016-ddos-trends-layer-7-ddos-attacks-a-growing-trend/}}.   (\bibinfo{date}{Aug} \bibinfo{year}{2016}).
\newblock


\bibitem[\protect\citeauthoryear{Wiki}{Wiki}{2023}]%
        {DVFS}
\bibfield{author}{\bibinfo{person}{Linux Wiki}.} \bibinfo{year}{2023}\natexlab{}.
\newblock \bibinfo{title}{{{CPU}} Frequency Scaling - {{ArchWiki}}}.
\newblock \bibinfo{howpublished}{\url{https://wiki.archlinux.org/title/CPU\_frequency\_scaling}}.   (\bibinfo{year}{2023}).
\newblock


\bibitem[\protect\citeauthoryear{Yin and Kaur}{Yin and Kaur}{2016}]%
        {yin2016can}
\bibfield{author}{\bibinfo{person}{Qianwen Yin} {and} \bibinfo{person}{Jasleen Kaur}.} \bibinfo{year}{2016}\natexlab{}.
\newblock \showarticletitle{Can machine learning benefit bandwidth estimation at ultra-high speeds?}. In \bibinfo{booktitle}{{\em International Conference on Passive and Active Network Measurement}}. Springer, \bibinfo{publisher}{Springer}, \bibinfo{pages}{397--411}.
\newblock


\end{thebibliography}

\end{document}